\documentclass[prl, amsmath, onecolumn, amssymb, aps, superscriptaddress, floatfix, preprintnumbers, notitlepage, reprint]{revtex4-1}
\usepackage{times}
\usepackage{graphicx}
\usepackage{amsmath, braket, amsfonts}
\usepackage{amssymb}
\usepackage{natbib}
\usepackage{todonotes}
\usepackage[modulo]{lineno}
\usepackage{bm, color, ulem}

\newcommand{\gi}{\lambda_{\textrm{I}}}
\newcommand{\gr}{\lambda_{\textrm{R}}}

\newcommand{\be}{\begin{equation}}
\newcommand{\ee}{\end{equation}}

\begin{document}

\title{Spin-orbit driven band inversion in bilayer graphene by van der Waals proximity effect}
\author{J.O. Island}
\affiliation{Department of Physics, University of California, Santa Barbara CA 93106 USA}
\author{X. Cui}
\affiliation{Department of Physics, University of California, Santa Barbara CA 93106 USA}
\author{C. Lewandowski}
\affiliation{Department of Physics, Massachusetts Institute of Technology, 77 Massachusetts Avenue, Cambridge, Massachusetts 02139, USA}
\author{J.Y. Khoo}
\affiliation{Department of Physics, Massachusetts Institute of Technology, 77 Massachusetts Avenue, Cambridge, Massachusetts 02139, USA}
\author{E. M. Spanton}
\affiliation{Department of Physics, University of California, Santa Barbara CA 93106 USA}
\author{H. Zhou}
\affiliation{Department of Physics, University of California, Santa Barbara CA 93106 USA}
\author{D. Rhodes}
\affiliation{Department of Mechanical Engineering, Columbia University, New York, NY 10027, USA}
\author{J.C. Hone}
\affiliation{Department of Mechanical Engineering, Columbia University, New York, NY 10027, USA}
\author{T. Taniguchi}
\affiliation{National Institute for Materials Science, Tsukuba, Ibaraki 305-0044, Japan}
\author{K. Watanabe}
\affiliation{National Institute for Materials Science, Tsukuba, Ibaraki 305-0044, Japan}
\author{L.S. Levitov}
\affiliation{Department of Physics, Massachusetts Institute of Technology, 77 Massachusetts Avenue, Cambridge, Massachusetts 02139, USA}
\author{M.P. Zaletel}
\affiliation{Department of Physics, University of California, Berkeley, CA 94720 USA}
\author{A.F. Young}
\affiliation{Department of Physics, University of California, Santa Barbara CA 93106 USA}

\begin{abstract}
Spin orbit coupling (SOC) is the key to realizing time-reversal invariant topological phases of matter
\cite{hasan_colloquium:_2010,qi_topological_2011}. Famously, SOC was predicted by Kane and Mele\cite{kane_quantum_2005} to stabilize a quantum spin Hall insulator; however, the weak intrinsic SOC in monolayer graphene\cite{min_intrinsic_2006,huertas-hernando_spin-orbit_2006,yao_spin-orbit_2007,sichau_intrinsic_2017} has precluded experimental observation. Here, we exploit a layer-selective proximity effect---achieved via van der Waals contact to a semiconducting transition metal dichalcogenide\cite{gmitra_graphene_2015,gmitra_proximity_2017,wang_strong_2015,wang_origin_2016,yang_tunable_2016,yang_strong_2017,volkl_magnetotransport_2017, wakamura_strong_2018, zihlmann_large_2018,avsar_spinorbit_2014, dankert_electrical_2017, ghiasi_large_2017, omar_spin_2018, benitez_strongly_2018}---to engineer Kane-Mele SOC in ultra-clean \textit{bilayer} graphene.
Using high-resolution capacitance measurements to probe the bulk electronic compressibility, we find that SOC leads to the formation of a distinct incompressible, gapped phase at charge neutrality.
The experimental data agrees quantitatively with a simple theoretical model in which the new phase results from SOC-driven band inversion.
In contrast to Kane-Mele SOC in monolayer graphene, the inverted phase is not expected to be a time reversal invariant topological insulator, despite being separated from conventional band insulators by electric field tuned phase transitions where crystal symmetry mandates that the bulk gap must close\cite{zaletel_gate-tunable_2019}.
Electrical transport measurements, conspicuously, reveal that the inverted phase has a conductivity $\sim e^2/h$, which is suppressed by exceptionally small in-plane magnetic fields.  The high conductivity and anomalous magnetoresistance are consistent with theoretical models that predict helical edge states within the inversted phase, that are protected from backscattering by an emergent spin symmetry that remains robust even for large Rashba SOC.
Our results pave the way for proximity engineering of strong topological insulators as well as correlated quantum phases in the strong spin-orbit regime in graphene heterostructures.
\end{abstract}
\maketitle

Depending on microscopic symmetry, spin orbit coupling (SOC) in graphene can take several forms, leading in turn to different electronic states at charge neutrality. In the absence of SOC, the low energy electronic structure of monolayer graphene is described by Dirac equations in two inequivalent valleys centered at the two momenta $K$ and $K'$ of the hexagonal Brillouin zone.
SOC, along with other perturbations that break the equivalence of the two valleys or two carbon sublattices, can be written with the aid of three sets of Pauli matrices, $\hat \sigma_i$, $\hat \tau_i$ and $\hat s_i$, which operate on the space of the carbon sublattices within the graphene unit cell, the K and K' valleys, and the physical electron spin, respectively. If the full symmetry of the graphene crystal is preserved, the only symmetry-allowed SOC term at low energies is the Kane-Mele (KM) term\cite{kane_quantum_2005}, $H_{\text{KM}}=\frac{\lambda_{\text{KM}}}{2} \sigma_z\tau_z s_z$. However, additional terms can arise when experimental substrates break lattice symmetries; these include the Rashba SOC, $H_\text{R}=\lambda_\text{R}\left(\sigma_x\tau_z s_y-\sigma_y s_x\right)$, and the so-called Ising SOC, $H_\text{I}=\lambda_\text{I} \tau_z s_z$.
In monolayer graphene, only the intrinsic $\lambda_{\text{KM}}$ term leads to a topological phase\cite{kane_quantum_2005,yang_tunable_2016}. Moreover, this topological phase requires $\lambda_\text{R}<\lambda_{\text{KM}}$\cite{kane_quantum_2005}, a physically unrealistic\cite{min_intrinsic_2006, huertas-hernando_spin-orbit_2006,yao_spin-orbit_2007} scenario given the measured value $\lambda_{\text{KM}}\approx$40 $\mu$eV\cite{sichau_intrinsic_2017}.

\begin{figure*}[t!]
\includegraphics[width=3.5 in]{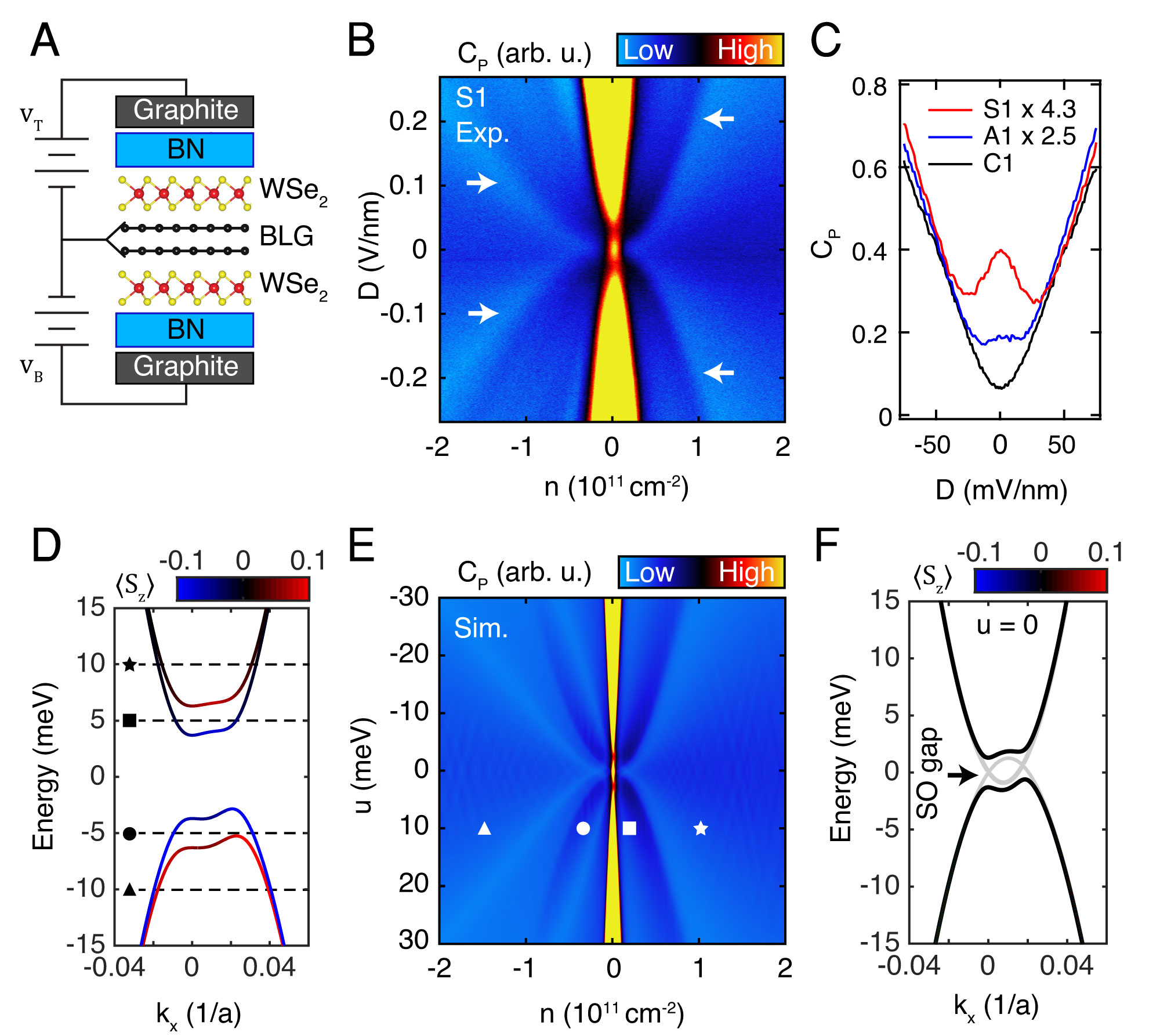}
\caption{\textbf{Inverted phase in bilayer graphene from proximity induced SOC.}
(A) Device schematic for a symmetrically WSe$_2$ encapsulated device. The charge density $n=c_t v_t+c_b v_b$ and perpendicular displacement field $D=\frac{1}{\epsilon_0}\left(c_t v_t-c_b v_b\right)$ are controlled by the voltages applied to the top and bottom gate ($v_{t(b)}$).
(B) Penetration field capacitance, $C_P$, as a function of charge density $n$ and displacement field $D$ measured at B=0 and T$\approx$50 mK in the Device S1. White arrows indicate compressibility minima associated with band splitting due to proximity induced SOC.
(C) $C_P$ measured at $n=0$ for a device symmetrically encapsulated with WSe$_2$ (S1, red), a device asymmetrically encapsulated in hBN and WSe$_2$ (A1, blue) and a control device fully encapsulated in hBN (C1, black). Only the symmetrically encapsulated device shows an incompressible peak at $D=0$.
(D) Low energy bands near the $K$ point of the Brillouin zone for $u=10$ meV, calculated for a model that includes an Ising SOC of equal magnitude ($\lambda_\text{I}=2.6$ meV) but opposite signs on each layer.
(E) Simulated $C_P$ from for the same model. The symbols denote points corresponding to different Fermi levels called out in (D).
(F) Low energy bands near the $K$ point of the Brillouin zone for $u=0$, calculated within the same model. The bands are spin degenerate, and a gap is visible near zero energy.  Light gray bands are calculated in the absence of SOC.  
}
\label{fig1}
\end{figure*}

Proximity effects between two dimensional crystals provide a tool for engineering electronic structures that do not occur naturally within a single material. First principles calculations indicate \cite{gmitra_graphene_2015,gmitra_proximity_2017} that heterostructures of graphene and transition metal dichalcogenide semiconductors such as tungsten diselenide (WSe$_2$) may endow graphene electrons with a SOC strength of several meV---two orders of magnitude larger than the intrinsic KM SOC\cite{sichau_intrinsic_2017} and sufficient, in principle, to enable observation of new topological phases\cite{gmitra_trivial_2016}.
Numerous experimental efforts have reported signatures of enhanced SOC in graphene-transition metal dichalcogenide (TMD) heterostructures. However, most rely on measurements of either spin relaxation\cite{avsar_spinorbit_2014, dankert_electrical_2017, ghiasi_large_2017, omar_spin_2018, benitez_strongly_2018} or weak antilocalization\cite{wang_strong_2015,wang_origin_2016,yang_tunable_2016,yang_strong_2017,volkl_magnetotransport_2017, wakamura_strong_2018, zihlmann_large_2018}, neither of which can distinguish between bulk or defect mediated SOC.  One study\cite{wang_origin_2016} of quantum oscillations found evidence for Rashba SOC within graphene band electrons of BLG, but neither Ising SOC nor a $\lambda_{\textrm{KM}}$-driven topological phase\cite{kane_quantum_2005} have been reported.

Here we explore the effects of proximity-induced SOC van der Waals heterostructures built around bilayer graphene (BLG)-tungsten diselenide (WSe$_2$) interfaces (see Materials and Methods). In constrast to monolayer graphene, the electronic spectrum of BLG features a quadratic band touching at charge neutrality.  An applied perpendicular electric displacement field, $D$, drives the system to a layer polarized band insulator in which wavefunctions are strongly polarized on the low energy layer, making BLG an ideal tool for probing short-range van der Waals proximity effects.
To access subtle features within the electronic structure, we use hexagonal boron nitride (hBN) gate dielectrics\cite{dean_boron_2010} and single crystal graphite (G) top- and bottom-gate electrodes, which reduce charge inhomogeneity\cite{zibrov_tunable_2017} while simultaneously allowing independent control over the charge carrier density, $n$, and $D$ (see Fig. 1A). We measure the penetration field capacitance, $C_P$, which is inversely related to the bulk electronic compressibility\cite{SI}. Fig. 1B shows $C_P$ as a function of $n$ and $D$ measured in a BLG flake symmetrically encapsulated in WSe$_2$. The most prominent features of the data are incompressible states at $n \sim 0$ associated with layer-polarized band insulators (BIs), which deepen as $|D|$ increases and are characteristic of BLG. However, WSe$_2$ encapsulation also produces features associated with proximity-induced SOC.
First, four weak $C_P$ minima appear at finite $n$ (indicated by arrows in Fig. 1B), whose positions depend strongly on $D$. Second, an additional incompressible phase is observed at charge neutrality near $D=0$, separated from the BIs by points of high compressibility. We observe the incompressible state in two devices, S1 and S2, that are symmetrically encapsulated in WSe$_2$ but it does not appear in either hBN-encapsulated device C1 or in devices A1 and A2 that are asymmetrically encapsulated in hBN and WSe$_2$ (Figs. 1C and \ref{figs1d}).

Both experimental features can be captured by a continuum model of BLG\cite{SI} with the sole addition of an Ising SOC term having equal magnitude---but opposite sign---on the two carbon layers (Fig. 1D-F). The opposite sign is consistent with  3D inversion symmetry. Fig. 1E shows simulated $C_P$ from this model, in which the only free parameters are the strength of the SOC ($\lambda_\text{I}=2.6$~meV) and the out-of-plane dielectric constant of the BLG ($\epsilon^\perp_{\text{BLG}}=4.3$). The latter is used to relate the experimentally measured $D$ to the the interlayer potential difference $u=-\frac{d}{\epsilon^\perp_{\text{BLG}}}D$ which enters the theoretical model (here $d=.33$~nm is the BLG interlayer separation).
At finite $u$, Ising SOC splits the normally spin-degenerate conduction and valence bands in bilayer graphene (Fig. 1D).  The $C_P$ minima arise from the high compressibility associated with the edge of the higher (lower) conduction (valence) band.  These band edges appear at finite density after the first bands have already begun filling (Fig. 1D-E).

The $C_P$ maximum observed near $n,D=0$ can be understood by noting that within a two-band model of BLG, layer and carbon sublattice are equivalent\cite{mccann_landau-level_2006}. As a result, Ising SOC with opposite signs on opposite layers is equivalent to a single SOC term proportional to $\sigma_z$, precisely reproducing the Kane-Mele SOC\cite{kane_quantum_2005}. The two-band Hamiltonian of WSe$_2$ encapsulated BLG is thus
\begin{equation}
\hat H=\hat H_{\text{BLG}}+\frac{\lambda_\text{I}}{2} \sigma_z s_z\tau_z+\frac{u}{2} \sigma_z,
\end{equation}
where $\hat H_{\text{BLG}}$ describes BLG in the absence of either SOC or electric fields, and $\sigma^z$ indexes the two low energy carbon atoms, or equivalently, layer.

 Within this model, the SOC inverts the bands for $|u|<|\lambda_\text{I}|$, opening a gap even at $u=0$ (Fig. 1F). Due to the $2\pi$-Berry phase of the $\hat H_{\text{BLG}}$, however, the resulting inverted phase (IP) is not predicted to be a strong topological insulator and the observed incompressible phase near $n,D=0$ is not expected to have edge states protected by time-reversal symmetry. The IP is nevertheless topologically distinct from the high-$u$ BIs within this model; they differ in the polarization of the insulators' Wannier orbitals, which are pinned to one of three high-symmetry positions within the BLG unit cell\cite{zaletel_gate-tunable_2019,SI}. Theoretically, this distinction guarantees a gap closing between the IP and BIs, consistent with the compressible $C_P$ minima experimentally.

\begin{figure*}[t!]
\includegraphics[width=3.5 in]{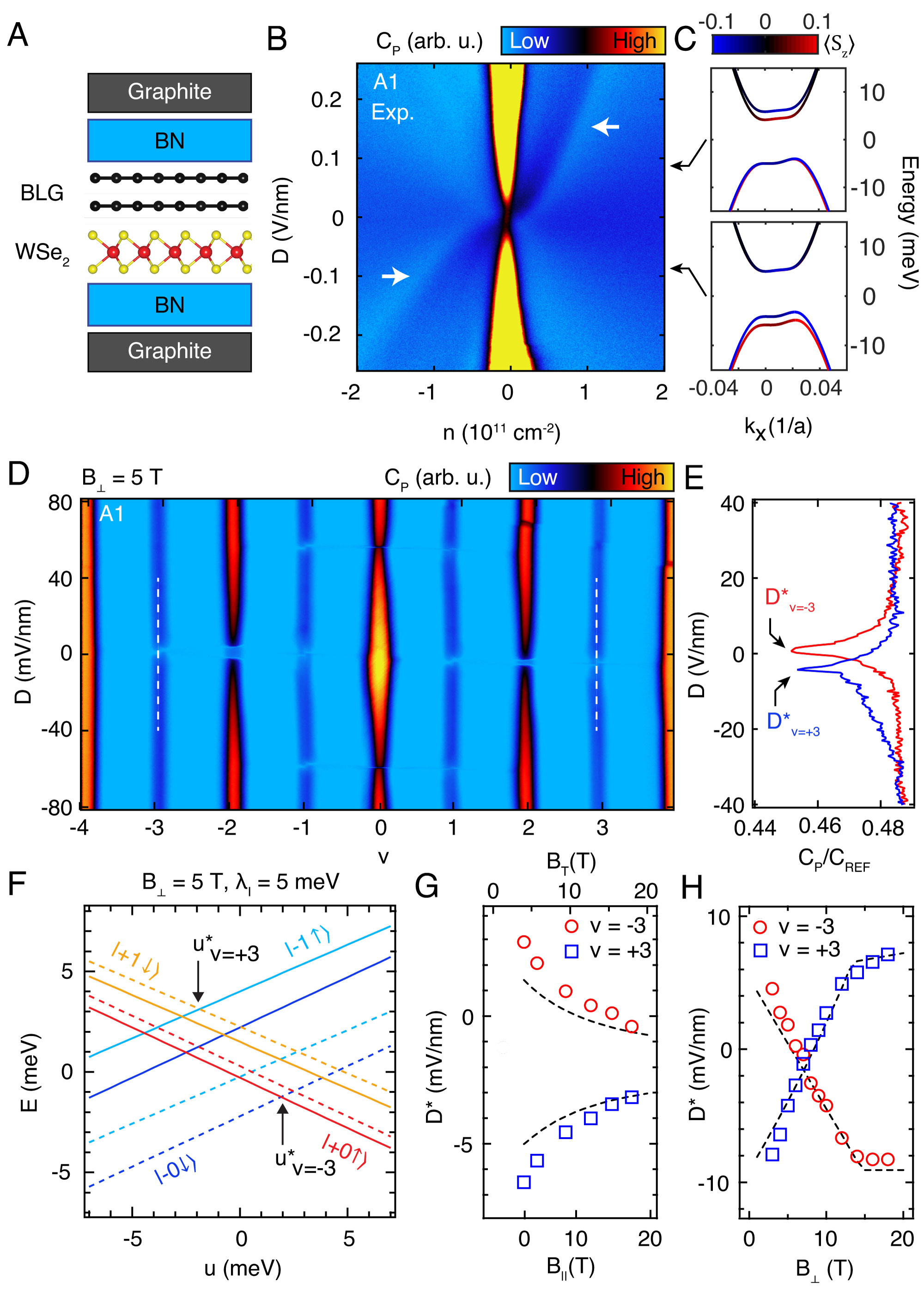}
\caption{\textbf{Layer selective spin-orbit proximity effect.}
(A) Schematic of an asymmetrically encapsulated device.
(B) Penetration field capacitance, $C_P$, as a function of $n$ and $D$ in the Device A1, measured at B=0 and T$\approx$50 mK. The white arrows indicate compressibility minima associated with band splitting, which we attribute to proximity induced SOC.
(C) Low energy bands near the $K$ point of the Brillouin zone for $u=-10$ meV (top) and $u=10$ meV (bottom), calculated for a model that includes an Ising SOC of ($\lambda_\text{I}=1.7$ meV) on the bottom layer.
(D) $C_P$ as a function of $\nu=2\pi\ell_B^2 n$ and $D$ for device A1 at $B_\perp=5$~T.
(E) $C_P$ traces taken at the locations of the white dotted lines in panel (D) for $\nu=\pm3$. The minima in $C_P$ correspond to phase transitions that occur at $D^*_{\nu=\pm3}$.
(F) Energy spectrum of the octet LL calculated within a continuum model with $\lambda_\text{I}=5$~meV on the bottom layer. The energy levels are plotted as a function of $u$ for $B_\perp=5$~T. Solid/dashed line denote spin projected parallel/antiparallel to the applied $B_\perp$. The indicated $u^*_{\nu=\pm3}$ correspond to the layer transitions for filling or emptying a single level from the octet. Note that for the chosen sign of $\lambda_\text{I}$, the net spin splitting on the proximity-affected layer is reversed.
(G) Measured $D^*_{\nu=\pm3}$---proportional to $u^*_{\nu=\pm3}$---as a function of $B_T$ for fixed $B_\perp=4$~T in device A1. The dashed lines are fits to our model \cite{SI} with $\lambda_\text{I}=1.7$~meV on the the bottom layer.
(H) Measured $D^*_{\nu=\pm3}$ as a function of $B_\perp$ with $B_\parallel=0$ in device A1. Dashed lines are fits to the same model as in panel (G).
}
\label{fig2}
\end{figure*}

Proximity induced SOC arises from overlaps between atomic orbitals, so is only expected to occur for the BLG layer in direct contact with the WSe$_2$.
Fig.~2A shows a schematic representation of an asymmetric heterostructure in which WSe$_2$ is in contact with only the bottom layer.
In contrast to the symmetric devices, $C_P$-minima appear in device A1 only for electrons for $D<0$ and only for holes for $D>0$ (Fig. 2B). Note that in isolated BLG, all thermodynamic features should  respect $D\leftrightarrow -D$ symmetry, suggesting that the asymmetric $C_P$ minima are caused by coupling between the BLG and other elements in the heterostructure.  Theoretical simulations (Figs. 2C and \ref{figs1d}) confirm that when SOC is induced only on the layer proximate to the WSe$_2$, the spin splitting is restricted to the conduction (for $D>0$) or valence (for $D<0$), consistent with the experimental data\cite{SI}.

The nature of the induced SOC can be precisely validated using the unique properties of BLG in a quantizing perpendicular magnetic field ($B_\perp$), where the energy spectrum is described by highly degenerate Landau Levels (LLs). Twofold quasi-degeneracies of the inequivalent valleys ($\xi=\pm1$), spin projections ($s=\pm1$), and the lowest two orbital ($N=0,1$ ) LLs\cite{mccann_landau-level_2006} combine to form an octet LL spanning $\nu\in (-4,4)$, where  $\nu=2\pi\ell_B^2 n$ denotes the LL filling factor.  The degeneracy between individual octet sublevels $|\xi N s\rangle$ is lifted by small intrinsic level splittings that include the Zeeman effect (lifting the spin degeneracy), interlayer potential $u$ (lifting the valley degeneracy through the near-perfect equivalence of valley and layer polarization in the lowest LLs \cite{mccann_landau-level_2006}) and band structure effects that distinguish the two orbitals. If disorder is sufficiently low, these splittings tend to fully polarize the electron system into one or more of the $|\xi N s\rangle$, which manifests experimentally as incompressible phases at all integer $\nu$. As shown in Fig. 2D, this signature is present in WSe$_2$ supported samples, which show signs of full lifting of the octet degeneracy for $B_\perp\gtrsim 2$~T.
Notably, the higher $B_\perp$ phenomenology of our samples suggests that WSe$_2$ is comparable to hBN\cite{dean_boron_2010} as a substrate for high quality graphene heterostructures, with Coulomb driven states with fractional Hall conductivity observed in compressibility (see. Fig. \ref{figS5}-\ref{figS4}).

Three features make the octets LL a precision probe of SOC. First, the small scale of the intrinsic splittings allows even few-meV scale SOC\cite{gmitra_graphene_2015} to rearrange the LL filling sequence\cite{khoo_tunable_2018}. Second, the octet LL is entirely insensitive to Rashba SOC\cite{khoo_tunable_2018}, allowing a direct measurement of the Ising SOC\cite{khoo_tunable_2018}. Finally, a broad set of $D$-field tuned phase transitions are observed throughout the zero-energy LL, corresponding to transitions between states with differing layer polarizations. The critical displacement field, $D^*$, required to effect these transitions (Fig. 2E) provides a direct comparison of the energy for LLs on opposite layers (and thus opposite valleys), and can be extracted with high precision (Fig. 2E).
Fig. 2F shows the single particle energy spectrum of the octet LL with asymmetric Ising SOC\cite{SI}. While the Coulomb interaction changes the order in which these levels fill for $-3<\nu<3$, it plays no role in determining which LL fills first ($\nu = -3$) or last ($\nu = 3$)\cite{hunt_direct_2017}. We thus focus on the observed behavior of $D^*_{\nu=\pm3}$, which can be simply related to $u^*_{\nu=\pm3}$ calculated from our theoretical model\cite{SI}.  The spin structure of the LLs is readily probed by varying the in-plane magnetic field $B_\parallel$ at fixed $B_\perp$, which varies the Zeeman energy but leaves orbital energy scales fixed.  In hBN encapsulated devices, $D^*_{\nu=\pm3}$ is observed to be independent from $B_\parallel$ (Fig. \ref{figS2_3}), consistent with theoretical expectation that the transition occurs between ground states of  identical spin. A strikingly different dependence is observed in device A1 (Fig. 2D); now $D^*_{\nu=\pm3}$ is a strong function of $B_\parallel$ indicating that the transition is between ground states with different spin.
As shown in the figure, the observed behavior is quantitatively consistent with our model of Ising SOC on the WSe$_2$ proximal layer, under the stipulations that $\lambda_\text{I}$ be larger in magnitude than the Zeeman energy due to the applied magnetic field ($\lambda_\text{I}>E_Z$) and that the sign of $\lambda_\text{I}$ be chosen to cancel, rather then add, to the $B_{\perp}$-proportioal part of the Zeeman splitting in the affected valley.  
Under these conditions, the level structure near $\nu=\pm3$ is inverted, leading to the observed behavior in $B_\parallel$ (see Fig. \ref{figS_LLs}).  Moreover, $\lambda_\text{I}$ can now be directly extracted from the dependence of $D^*_{\nu=\pm3}$ on $B_\perp$\cite{khoo_tunable_2018}, in the absence of $B_\parallel$: because the effective Zeeman splitting arising from the SOC is oriented out of plane, an out-of-plane extrinsic Zeeman splitting will precisely cancel it when $2E_Z=2g\mu_B B_\perp=\lambda_\text{I}$. Fig. 2E shows $D^*_{\nu\pm3}$ for $B_\parallel=0$. The two curves cross at $B_\perp\approx 7.4$~T from which it follows that $\lambda_\text{I}=1.7$ meV (see Fig. \ref{figS_A2S2} for a similar analysis in device A2, resulting in $\lambda_\text{I}=2.0$~meV).  This value is in reasonable agreement with \textit{ab initio} calculations\cite{gmitra_trivial_2016} predicting $\lambda_\text{I}=1.19$ meV for similar device geometries. 

\begin{figure*}[t!]
\includegraphics[width=4.5 in]{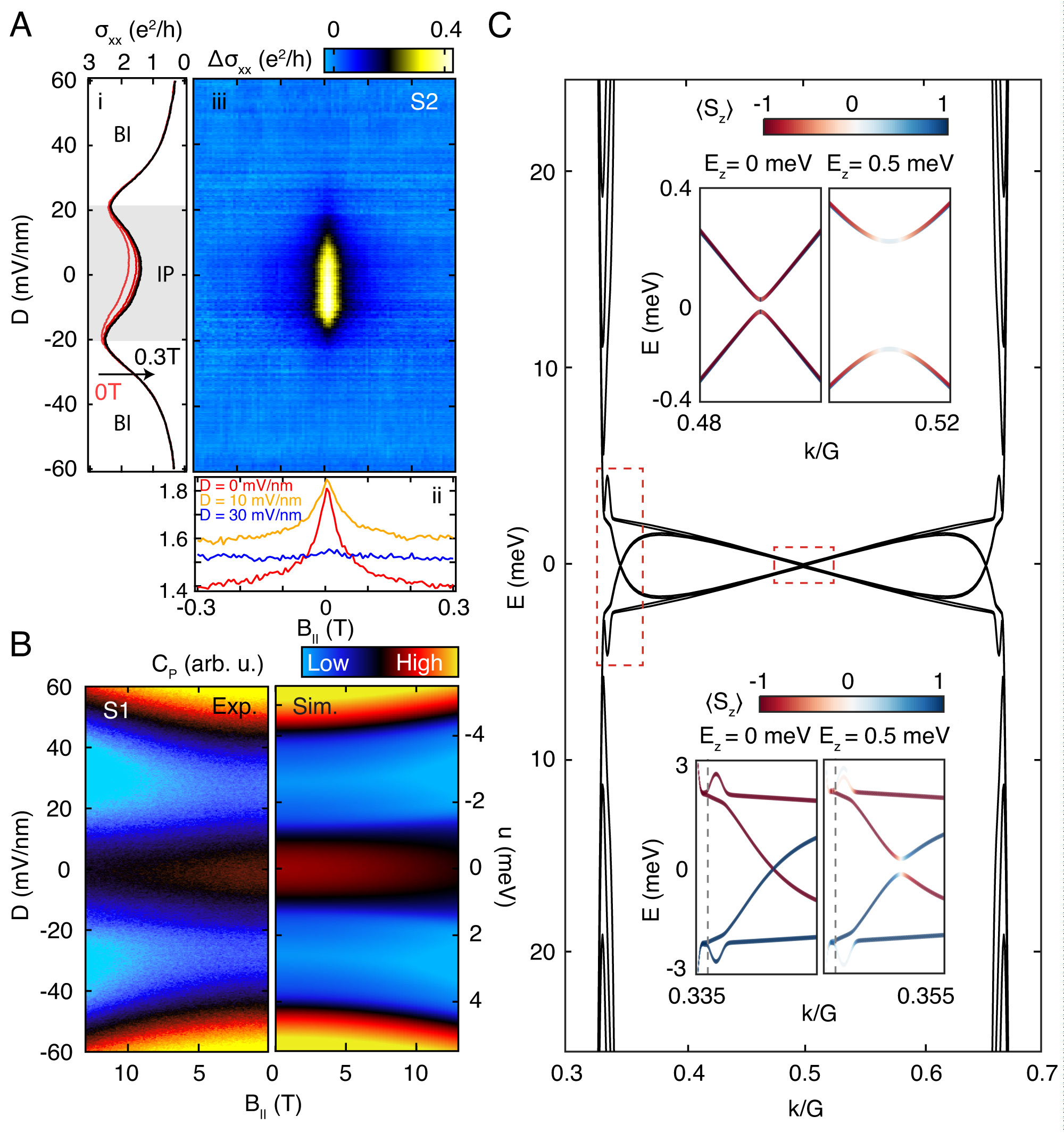}
\caption{\textbf{Magnetoconductance and edge state transport in the inverted phase.}
(A) Longitudinal conductance, $\sigma_{xx}=1/R_{xx}$, measured in a four terminal geometry at $n=0$.  (i) $\sigma_{xx}$ as a function of $D$ for different values of $B_\parallel$.  The IP is visible as a conductance suppression between $|D|<20$meV/nm, separated by conductance maxima from the BIs at large $|D|$. (ii) $\sigma_{xx}$ as a function of $B_\parallel$ for different values of $D$. Strong in-plane magnetoconductance is observed only in the IP, and not in the BIs.
(iii) Subtracted $\Delta\sigma_{xx}=\sigma_{xx}(B_\parallel)-\sigma_{xx}(B_\parallel=300 \text{mT})$, highlighting that the anomalous in-plane magnetoconductance is restricted to the IP.
(B) $C_P$ as a function of $D$ and $B_\parallel$ with $B_\perp=0$~T. Experimental data for device S1 is plotted on the left and simulated $C_P$ from the continuum model is plotted on the right.
(C) Band structure calculated for a 1000 lattice-site wide BLG ribbon with Ising SOC of equal magnitude ($\lambda_\text{I}=5$ meV) and opposite sign on the opposite layers.
Within the bulk gap, a set of spin polarized energy bands emerge with wave functions tightly localized on the sample boundary. The top inset shows a detailed view of the edge states that approach E=0 near the $M$ point of the Brillouin zone for $E_Z=g\mu_B B_\parallel=0$ (Left) and $E_Z=0.5$ meV (right). These states are not gapless even for $E_Z=\lambda_\text{R}=0$. The lower panel shows the same calculation for edge states near the $K$ point. While gapless for $E_Z=\lambda_\text{R}=0$, in-plane field produces a gap proportional to $E_Z$.
}
\label{fig3}
\end{figure*}

In summary, to account for the bulk thermodynamic measurements of TMD-encapsulated BLG it is necessary to include Ising SOC $\gi$ on the layers proximate to the TMD, with equal and opposite values in the case of symmetric-encapsulation. However, our bulk measurements do not quantitatively constrain the Rashba SOC $\gr$. This is because the effect of $\gr$ on the low-energy band structure of BLG is highly suppressed by the  large inter-layer hybridization scale $\gamma_1$, and hence has a negligible effect on both the LL structure\cite{khoo_tunable_2018} and the zero magnetic field compressibility\cite{zaletel_gate-tunable_2019}. Previous experimental works largely focusing on monolayer devices have suggested a wide range of values for $\gr$ between 0 and 15 meV
\cite{yang_strong_2017,wang_origin_2016,omar_spin_2018, volkl_magnetotransport_2017, garcia_spin_2018}.
Despite its small influence on the bulk, theoretically the Rashba SOC plays a critical role in the \textit{edge state} spectrum of our BLG Hamiltonian, Eq. 1. For $\lambda_\text{R}=0$, $s_z$ is conserved, and the edge is predicted to host  two pairs of counter-propagating, spin filtered modes with  a quantized conductance (and spin Hall effect) of $4 \frac{e^2}{h}$\cite{gmitra_trivial_2016,zaletel_gate-tunable_2019}. When $\lambda_\text{R}\neq0$, however, $s_z$ symmetry is weakly broken and backscattering is expected.

In order to explore the possibility of edge state transport experimentally, we measure the longitudinal resistance ($R_{xx}=1/\sigma_{xx}$) as a function of $D$ and $n$ at $B_T=0$ for a symmetrically encapsulated device (S2) fabricated with transport contacts. The IP  is readily evident (Figs. 3A(i) and \ref{figS_A2S2}B), but is not a strong insulator, showing a finite four-terminal conductivity $\sigma_{xx}\gtrsim e^2/h$ at low temperatures. Such behavior is consistent with transport via edge states with a finite transmission coefficient, as might be expected for a finite length helical edge in which backscattering is suppressed but not completely forbidden. However, it is also consistent with bulk conduction through a mesoscopically disordered sample.

To disambiguate these two scenarios, we study the response of $\sigma_{xx}$ to a small in-plane magnetic field,  $B_\parallel$, which breaks $\sigma_z$ conservation. $B_\parallel$  is expected to rapidly localize spin-filtered edge states while having a minimal effect on the bulk energy spectrum when $E_Z<\lambda_\text{I}$ (i.e. $B_\parallel<10-20$~T for $\lambda_\text{I} \sim$ 1-2meV). Indeed, the IP  shows strong in-plane magneto-resistance at exceptionally small values of $B_\parallel$, as shown in Figs. 3A(i-iii), a response that is completely absent in the BI phases. This is in contrast with $C_P$ measurements of the bulk, which show no detectable dependence on $B_\parallel$ at low fields and a high field response that is well accounted for in our theoretical model (Fig. 3B). 

To explore the plausibility of edge state transport in the IP, we perform numerical simulations of the band structure 
for a finite width BLG ribbon\cite{SI}. The resulting energy levels (Fig. 3C) reproduce the bulk gap from the continuum model, but additionally feature states localized on the edge of the ribbon that approach $E=0$ at the $K$, $K'$, and $M$ points in the Brillouin zone. Closer inspection reveals that while the $M$ point (Fig. 3C, top inset) edge states are gapped even in the absence of Rashba SOC, the states near the K and K' points (Fig. 3C, bottom inset) are indeed gapless and helical. Simulations\cite{zaletel_gate-tunable_2019} show that the helical edge states are remarkably robust to Rashba SOC: the edge gap is strongly suppressed by interlayer hopping terms in the BLG band structure, so that the anticipated gap is $\Delta_{edge}\propto \lambda_\text{R}^2/\gamma_1\approx.25 $~meV even for $\lambda_\text{R}=10 $~meV, near the largest values reported in the literature\cite{wang_origin_2016}. In contrast, the helical edge states are highly sensitive to finite $B_\parallel$ (Fig. 3C, right panels of top and bottom insets), developing an energy gap directly proportional to the Zeeman energy $\Delta_{edge}\approx g\mu_BB_\parallel\approx .1$~meV $\times B_\parallel/\text{T}$.

These considerations are qualitatively consistent with the observed in-plane magnetoresistance anomaly. However, we note that the fine structure of the edge states is highly sensitive to choice of theoretical parameters, and ignores electron-electron interactions which may play a significant role. Our experimental magnetoresistance data saturate at a rather low resistance value, well below $h/e^2$, possibly indicating residual bulk conductance.
It bears noting that the weak edge transport is understood to be a consequence of the accidental approximate $s_z$-conservation in this system. However, recent theoretical work\cite{zaletel_gate-tunable_2019} argues that the same fabrication technique implemented with ABC trilayer graphene---where the SOC gaps a cubic band touching with  $3\pi$ Berry phase---results in a strong topological insulator with time-reversal symmetry protected edge states.
Notably, the multi-layer graphene/WSe$_2$ platform generically allows for gate tunable transitions between topological and trivial insulating states, a long-standing milestone in the quest for reconfigurable topological circuits.

\paragraph{Acknowledgments: }
Experimental work at UCSB was supported by the ARO under award MURI W911NF-16-1-0361.
DR, JH acknowledge support by the U.S. Department of Energy, DE-SC0016703 for synthesis of WSe$_2$ crystals.
KW and TT acknowledge support from the Elemental Strategy Initiative conducted by the MEXT, Japan and and the CREST (JPMJCR15F3), JST.
MZ was supported by the Director, Office of Science, Office of Basic Energy Sciences, Materials Sciences and Engineering Division of the U.S. Department of Energy under contract no. DE-AC02-05-CH11231 (van der Waals heterostructures program, KCWF16)AFY acknowledges the support of the David and Lucile Packard Foundation and and Alfred. P. Sloan Foundation.
JOI acknowledges the support of the Netherlands Organization for Scientiï¬c Research (NWO) through the Rubicon grant, project number 680-50-1225/2474.
JYK acknowledges support by the National Science Scholarship from the Agency for Science, Technology and Research (A*STAR).
A portion of this work was performed at the National High Magnetic Field Laboratory, which is supported by National Science Foundation Cooperative Agreement No. DMR-1644779 and the State of Florida.
Measurements made use of a dilution refrigerator funded through the Major Research Instrumentation program of the U.S. National Science Foundation under award no. DMR-1531389, and the MRL Shared Experimental Facilities, which are supported by the MRSEC Program of the U.S. National Science Foundation under Award No. DMR-1720256.

\paragraph{Author Contributions:}

JOI, XC, and HZ fabricated the devices. JOI, XC, EMS, and AFY performed the measurements. JYK, CL, and MZ performed the theoretical simulations. JOI, XC, JYK, CL, MZ and AFY analyzed the data. DR and JH grew the WSe$_2$ crystals used in devices A1, S1, and A2/S2. TT and KW grew the hBN crystals used in all devices. JOI and AFY wrote the paper in consultation with XC, JYK, CL, MZ.

\paragraph{Materials and Methods: }
Five bilayer graphene van der Waals heterostructure devices (labeled C1, A1, S1, A2/S2, A3) were fabricated and studied in this work, as shown Fig. \ref{figS1a}. All devices were fabricated using a stacking and transfer method based on van der Waals adhesion\cite{wang_one-dimensional_2013}. Contact to the bilayer graphene was achieved using $\sim10$ nm thick graphite flakes. All devices made use of single crystal hexagonal boron nitride gate dielectrics\cite{dean_boron_2010} and graphite gates\cite{zibrov_tunable_2017}, which in combination are known to minimize extrinsic charge disorder. Devices A1, S1, and A2/S2 were fabricated using WSe$_2$ crystals grown by a flux method\cite{gustafsson_ambipolar_2018}, while device A3 was fabricated using WSe$_2$ from a commercial source (2Dsemiconductors.com).
A mixture of CHF$_3$ and O$_2$ was used to dry etch the stacks to define the device area and create connections to the bilayer graphene and top and bottom gates. Electrical contact was made to the edges of the exposed graphite flakes using a three layer metal film of Cr/Pd/Au (3nm/15nm/80nm).

Penetration field capacitance measurements were performed on all devices and additional connections were patterned in device A2/S2 in order to perform electrical transport. With the exception of device A2/S2, all devices had characteristics corresponding to uniform encapsulation on each facet by either hBN or WSe$_2$, depending on the device configuration. Device A2/S2 showed two sets of LL phase transitions in the zero-energy LL, consistent with the device having one asymmetric portion (A2) and one symmetric portion (S2). Extended data from this device is presented in Fig. \ref{figS_A2S2}.

Small changes in device capacitance are measured using a low temperature capacitance bridge\cite{ashoori_single-electron_1992} which effectively disconnects the device capacitance from the large capacitance of the cryostat cabling, see Fig. \ref{figS1b}. $C_P$ is a measure of the capacitance between the top and bottom gates and its magnitude is high when the bilayer graphene is incompressible (gapped) and low when it is compressible (conducting). $C_P$ is measured by applying a fixed AC excitation (17-33 kHz) to the top gate ($\delta V_{top}$) and the phase and amplitude of a second AC excitation with the same frequency is adjusted and applied to a standard reference capacitor ($C_{\text{ref}}$) on the low temperature amplifier in order to balance the capacitance bridge. A commercial high electron mobility transistor (FHX35X) transforms the small sample impedance to a 1 k$\Omega$ output impedance roughly translating to a (power) gain of $\sim$1000. $V_{top}$ and $V_{samp}$ (at a fixed $V_{gate}$) are swept in order to adjust charge density $n=c_Tv_T+c_Bv_B$ and displacement field $D=(c_T v_T-c_B v_B)/(2\epsilon_0)$.

All measurements were performed within the electronic band-gap of WSe$_2$. Charge accumulation in the WSe$_2$ layers is evident in capacitance measurements at high densities and manifests as apparent negative signals in $C_P$ as charge carriers are transferred from bilayer graphene to the opposite facet of the WSe$_2$ substrates.

\bibliographystyle{unsrt}

\let\oldaddcontentsline\addcontentsline
\renewcommand{\addcontentsline}[3]{}
\bibliography{references,supp}

\begin{thebibliography}{10}

\bibitem{hasan_colloquium:_2010}
M.~Z. Hasan and C.~L. Kane.
\newblock Colloquium: {Topological} insulators.
\newblock {\em Reviews of Modern Physics}, 82(4):3045--3067, 2010.

\bibitem{qi_topological_2011}
X.-L. Qi and S.-C. Zhang.
\newblock Topological insulators and superconductors.
\newblock {\em Reviews of Modern Physics}, 83(4):1057--1110, 2011.

\bibitem{kane_quantum_2005}
C.~L. Kane and E.~J. Mele.
\newblock Quantum {Spin} {Hall} {Effect} in {Graphene}.
\newblock {\em Phys. Rev. Lett.}, 95(22), November 2005.

\bibitem{min_intrinsic_2006}
Hongki Min, J.~E. Hill, N.~A. Sinitsyn, B.~R. Sahu, Leonard Kleinman, and A.~H.
  MacDonald.
\newblock Intrinsic and {Rashba} spin-orbit interactions in graphene sheets.
\newblock {\em Physical Review B}, 74(16):165310, October 2006.

\bibitem{huertas-hernando_spin-orbit_2006}
Daniel Huertas-Hernando, F.~Guinea, and Arne Brataas.
\newblock Spin-orbit coupling in curved graphene, fullerenes, nanotubes, and
  nanotube caps.
\newblock {\em Phys. Rev. B}, 74(15), October 2006.

\bibitem{yao_spin-orbit_2007}
Yugui Yao, Fei Ye, Xiao-Liang Qi, Shou-Cheng Zhang, and Zhong Fang.
\newblock Spin-orbit gap of graphene: {First}-principles calculations.
\newblock {\em Phys. Rev. B}, 75(4), January 2007.

\bibitem{sichau_intrinsic_2017}
Jonas Sichau, Marta Prada, Timothy~J. Lyon, Bojan Bosnjak, Tim Anlauf, Lars
  Tiemann, and Robert~H. Blick.
\newblock Intrinsic spin-orbit coupling gap and the evidence of a topological
  state in graphene.
\newblock {\em cond-mat}, September 2017.

\bibitem{gmitra_graphene_2015}
Martin Gmitra and Jaroslav Fabian.
\newblock Graphene on transition-metal dichalcogenides: {A} platform for
  proximity spin-orbit physics and optospintronics.
\newblock {\em Physical Review B}, 92(15):155403, October 2015.

\bibitem{gmitra_proximity_2017}
Martin Gmitra and Jaroslav Fabian.
\newblock Proximity {Effects} in {Bilayer} {Graphene} on {Monolayer} {WSe}2:
  {Field}-{Effect} {Spin} {Valley} {Locking}, {Spin}-{Orbit} {Valve}, and
  {Spin} {Transistor}.
\newblock {\em Physical Review Letters}, 119(14):146401, October 2017.

\bibitem{wang_strong_2015}
Zhe Wang, Dong-Keun Ki, Hua Chen, Helmuth Berger, Allan~H. MacDonald, and
  Alberto~F. Morpurgo.
\newblock Strong interface-induced spin–orbit interaction in graphene on
  {WS}2.
\newblock {\em Nature Communications}, 6:8339, September 2015.

\bibitem{wang_origin_2016}
Zhe Wang, Dong-Keun Ki, Jun~Yong Khoo, Diego Mauro, Helmuth Berger, Leonid~S.
  Levitov, and Alberto~F. Morpurgo.
\newblock Origin and {Magnitude} of `{Designer}' {Spin}-{Orbit} {Interaction}
  in {Graphene} on {Semiconducting} {Transition} {Metal} {Dichalcogenides}.
\newblock {\em Physical Review X}, 6(4):041020, October 2016.

\bibitem{yang_tunable_2016}
Bowen Yang, Min-Feng Tu, Jeongwoo Kim, Yong Wu, Hui Wang, Jason Alicea, Ruqian
  Wu, Marc Bockrath, and Jing Shi.
\newblock Tunable spin–orbit coupling and symmetry-protected edge states in
  graphene/{WS} 2.
\newblock {\em 2D Materials}, 3(3):031012, 2016.

\bibitem{yang_strong_2017}
Bowen Yang, Mark Lohmann, David Barroso, Ingrid Liao, Zhisheng Lin, Yawen Liu,
  Ludwig Bartels, Kenji Watanabe, Takashi Taniguchi, and Jing Shi.
\newblock Strong electron-hole symmetric {Rashba} spin-orbit coupling in
  graphene/monolayer transition metal dichalcogenide heterostructures.
\newblock {\em Physical Review B}, 96(4):041409, July 2017.

\bibitem{volkl_magnetotransport_2017}
Tobias Völkl, Tobias Rockinger, Martin Drienovsky, Kenji Watanabe, Takashi
  Taniguchi, Dieter Weiss, and Jonathan Eroms.
\newblock Magnetotransport in heterostructures of transition metal
  dichalcogenides and graphene.
\newblock {\em Physical Review B}, 96(12):125405, September 2017.

\bibitem{wakamura_strong_2018}
T.~Wakamura, F.~Reale, P.~Palczynski, S.~Guéron, C.~Mattevi, and H.~Bouchiat.
\newblock Strong {Anisotropic} {Spin}-{Orbit} {Interaction} {Induced} in
  {Graphene} by {Monolayer} \$\{{\textbackslash}mathrm\{{WS}\}\}\_\{2\}\$.
\newblock {\em Physical Review Letters}, 120(10):106802, March 2018.

\bibitem{zihlmann_large_2018}
Simon Zihlmann, Aron~W. Cummings, Jose~H. Garcia, Máté Kedves, Kenji
  Watanabe, Takashi Taniguchi, Christian Schönenberger, and Péter Makk.
\newblock Large spin relaxation anisotropy and valley-{Zeeman} spin-orbit
  coupling in {WSe}2/graphene/h-{BN} heterostructures.
\newblock {\em Physical Review B}, 97(7):075434, February 2018.

\bibitem{avsar_spinorbit_2014}
A.~Avsar, J.~Y. Tan, T.~Taychatanapat, J.~Balakrishnan, G.~K.~W. Koon, Y.~Yeo,
  J.~Lahiri, A.~Carvalho, A.~S. Rodin, E.~C.~T. O’Farrell, G.~Eda, A.~H.
  Castro~Neto, and B.~Özyilmaz.
\newblock Spin–orbit proximity effect in graphene.
\newblock {\em Nature Communications}, 5, September 2014.

\bibitem{dankert_electrical_2017}
André Dankert and Saroj~P. Dash.
\newblock Electrical gate control of spin current in van der {Waals}
  heterostructures at room temperature.
\newblock {\em Nature Communications}, 8:16093, July 2017.

\bibitem{ghiasi_large_2017}
Talieh~S. Ghiasi, Josep Ingla-Aynés, Alexey~A. Kaverzin, and Bart~J. van Wees.
\newblock Large {Proximity}-{Induced} {Spin} {Lifetime} {Anisotropy} in
  {Transition}-{Metal} {Dichalcogenide}/{Graphene} {Heterostructures}.
\newblock {\em Nano Letters}, 17(12):7528--7532, December 2017.

\bibitem{omar_spin_2018}
S.~Omar and B.~J. van Wees.
\newblock Spin transport in high-mobility graphene on {WS}2 substrate with
  electric-field tunable proximity spin-orbit interaction.
\newblock {\em Physical Review B}, 97(4):045414, January 2018.

\bibitem{benitez_strongly_2018}
L.~Antonio Benítez, Juan~F. Sierra, Williams~Savero Torres, Aloïs Arrighi,
  Frédéric Bonell, Marius~V. Costache, and Sergio~O. Valenzuela.
\newblock Strongly anisotropic spin relaxation in graphene–transition metal
  dichalcogenide heterostructures at room temperature.
\newblock {\em Nature Physics}, 14(3):303--308, March 2018.

\bibitem{zaletel_gate-tunable_2019}
Michael~P. Zaletel and Jun~Yong Khoo.
\newblock The gate-tunable strong and fragile topology of multilayer-graphene
  on a transition metal dichalcogenide.
\newblock {\em arXiv:1901.01294 [cond-mat]}, January 2019.
\newblock arXiv: 1901.01294.

\bibitem{gmitra_trivial_2016}
Martin Gmitra, Denis Kochan, Petra Hogl, and Jaroslav Fabian.
\newblock Trivial and inverted {Dirac} bands and the emergence of quantum spin
  {Hall} states in graphene on transition-metal dichalcogenides.
\newblock {\em Physical Review B}, 93(15):155104, April 2016.

\bibitem{dean_boron_2010}
C.~R. Dean, A.~F. Young, I.~Meric, C.~Lee, L.~Wang, S.~Sorgenfrei, K.~Watanabe,
  T.~Taniguchi, P.~Kim, K.~L. Shepard, and J.~Hone.
\newblock Boron nitride substrates for high-quality graphene electronics.
\newblock {\em Nature Nanotechnology}, 5:722--726, 2010.

\bibitem{zibrov_tunable_2017}
A.~A. Zibrov, C.~Kometter, H.~Zhou, E.~M. Spanton, T.~Taniguchi, K.~Watanabe,
  M.~P. Zaletel, and A.~F. Young.
\newblock Tunable interacting composite fermion phases in a half-filled
  bilayer-graphene {Landau} level.
\newblock {\em Nature}, 549(7672):360--364, September 2017.

\bibitem{SI}
Island et~al.
\newblock Supplementary material.
\newblock {\em Supplementary Material}, 2018.

\bibitem{mccann_landau-level_2006}
Edward McCann and Vladimir~I. Fal'ko.
\newblock Landau-{Level} {Degeneracy} and {Quantum} {Hall} {Effect} in a
  {Graphite} {Bilayer}.
\newblock {\em Phys. Rev. Lett.}, 96(8), March 2006.

\bibitem{khoo_tunable_2018}
Jun~Yong Khoo and Leonid Levitov.
\newblock Tunable quantum {Hall} edge conduction in bilayer graphene through
  spin-orbit interaction.
\newblock {\em Physical Review B}, 98(11):115307, September 2018.

\bibitem{hunt_direct_2017}
B.~M. Hunt, J.~I.~A. Li, A.~A. Zibrov, L.~Wang, T.~Taniguchi, K.~Watanabe,
  J.~Hone, C.~R. Dean, M.~Zaletel, R.~C. Ashoori, and A.~F. Young.
\newblock Direct measurement of discrete valley and orbital quantum numbers in
  bilayer graphene.
\newblock {\em Nature Communications}, 8(1):948, October 2017.

\bibitem{garcia_spin_2018}
Jose~H. Garcia, Marc Vila, Aron~W. Cummings, and Stephan Roche.
\newblock Spin transport in graphene/transition metal dichalcogenide
  heterostructures.
\newblock {\em Chemical Society Reviews}, 47(9):3359--3379, May 2018.

\bibitem{wang_one-dimensional_2013}
L.~Wang, I.~Meric, P.~Y. Huang, Q.~Gao, Y.~Gao, H.~Tran, T.~Taniguchi,
  K.~Watanabe, L.~M. Campos, D.~A. Muller, J.~Guo, P.~Kim, J.~Hone, K.~L.
  Shepard, and C.~R. Dean.
\newblock One-{Dimensional} {Electrical} {Contact} to a {Two}-{Dimensional}
  {Material}.
\newblock {\em Science}, 342(6158):614--617, November 2013.

\bibitem{gustafsson_ambipolar_2018}
Martin~V. Gustafsson, Matthew Yankowitz, Carlos Forsythe, Daniel Rhodes, Kenji
  Watanabe, Takashi Taniguchi, James Hone, Xiaoyang Zhu, and Cory~R. Dean.
\newblock Ambipolar {Landau} levels and strong band-selective carrier
  interactions in monolayer {WSe} 2.
\newblock {\em Nature Materials}, 17(5):411, May 2018.

\bibitem{ashoori_single-electron_1992}
R.~C. Ashoori, H.~L. Stormer, J.~S. Weiner, L.~N. Pfeiffer, S.~J. Pearton,
  K.~W. Baldwin, and K.~W. West.
\newblock Single-electron capacitance spectroscopy of discrete quantum levels.
\newblock {\em Phys. Rev. Lett.}, 68(20):3088--3091, May 1992.

\bibitem{spanton_observation_2018}
Eric~M. Spanton, Alexander~A. Zibrov, Haoxin Zhou, Takashi Taniguchi, Kenji
  Watanabe, Michael~P. Zaletel, and Andrea~F. Young.
\newblock Observation of fractional {Chern} insulators in a van der {Waals}
  heterostructure.
\newblock {\em Science}, 360(6384):62--66, April 2018.

\bibitem{jung_accurate_2014}
Jeil Jung and Allan~H. MacDonald.
\newblock Accurate tight-binding models for the pi bands of bilayer graphene.
\newblock {\em Physical Review B}, 89(3):035405, January 2014.

\bibitem{mccann_electronic_2013}
Edward McCann and Mikito Koshino.
\newblock The electronic properties of bilayer graphene.
\newblock {\em Reports on Progress in Physics}, 76(5):056503, 2013.

\end{thebibliography}
\let\addcontentsline\oldaddcontentsline


\newpage
\clearpage
\onecolumngrid
\let\oldaddcontentsline\addcontentsline
\renewcommand{\addcontentsline}[3]{}
\let\addcontentsline\oldaddcontentsline

\section{Supplementary materials}
\setcounter{equation}{0}
\setcounter{figure}{0}
\setcounter{table}{0}
\setcounter{page}{1}
\setcounter{section}{0}
\renewcommand{\thefigure}{S\arabic{figure}}
\renewcommand{\thesubsection}{S\arabic{subsection}}
\renewcommand{\theequation}{S\arabic{equation}}
\renewcommand{\thetable}{S\arabic{table}}
\renewcommand{\thepage}{S-\arabic{page}}

\let\oldaddcontentsline\addcontentsline
\renewcommand{\addcontentsline}[3]{}
\let\addcontentsline\oldaddcontentsline

\subsection{Device fabrication and measurement}

\begin{figure*}[h]
\includegraphics[width=\linewidth]{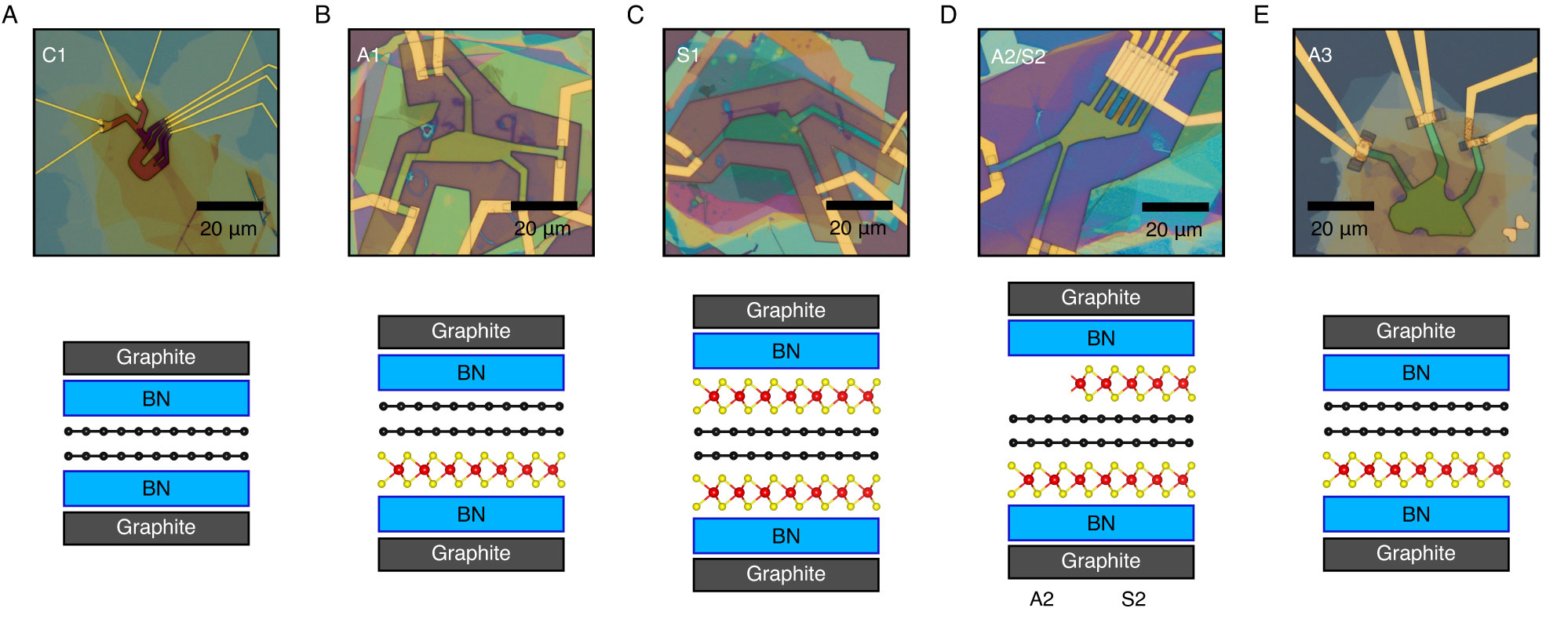}
\caption{
(A) Optical image of the control device C1 with accompanying model (below).
(B) Optical image of the asymmetric device A1 with accompanying model (below).
(C) Optical image of the symmetric device S1 with accompanying model (below).
(D) Optical image of another symmetric device with a single-sided region. Details about this device are presented in Fig. \ref{figS_A2S2}.
(E) Optical image of another asymmetric device A3 with accompanying model (below). This device showed additional features in the magnetocapacitance measurements which are associated with a moir\'e superlattice potential due to alignment of the bilayer graphene with the top BN, see Fig. \ref{figS4}.
}
\label{figS1a}
\end{figure*}

Five bilayer graphene van der Waals heterostructure devices (labeled C1, A1, S1, A2/S2, A3) were fabricated and studied in this work, as shown Fig. \ref{figS1a}. All devices were fabricated using a stacking and transfer method based on van der Waals adhesion\cite{wang_one-dimensional_2013}. Contact to the bilayer graphene was achieved using $\sim10$ nm thick graphite flakes. All devices made use of single crystal hexagonal boron nitride gate dielectrics\cite{dean_boron_2010} and graphite gates\cite{zibrov_tunable_2017}, which in combination are known to minimize extrinsic charge disorder. Devices A1, S1, and S2/S2 were fabricated using WSe$_2$ crystals grown by a flux method\cite{gustafsson_ambipolar_2018}, while device A3 was fabricated using WSe$_2$ from a commercial source (2Dsemiconductors.com).
A mixture of CHF$_3$ and O$_2$ was used to dry etch the stacks to define the device area and create connections to the bilayer graphene and top and bottom gates. Electrical contact was made to the edges of the exposed graphite flakes using a three layer metal film of Cr/Pd/Au (3nm/15nm/80nm).

Penetration field capacitance measurements were performed on all devices and additional connections were patterned in device A2/S2 in order to perform electrical transport. With the exception of device A2/S2, all devices had characteristics corresponding to uniform encapsulation on each facet by either hBN or WSe$_2$, depending on the device configuration. Device A2/S2 showed two sets of LL phase transitions in the zero-energy LL, consistent with the device having one asymmetric portion (A2) and one symmetric portion (S2). Extended data from this device is presented in Fig. \ref{figS_A2S2}.

\begin{figure*}[t!]
\includegraphics[width=5 in]{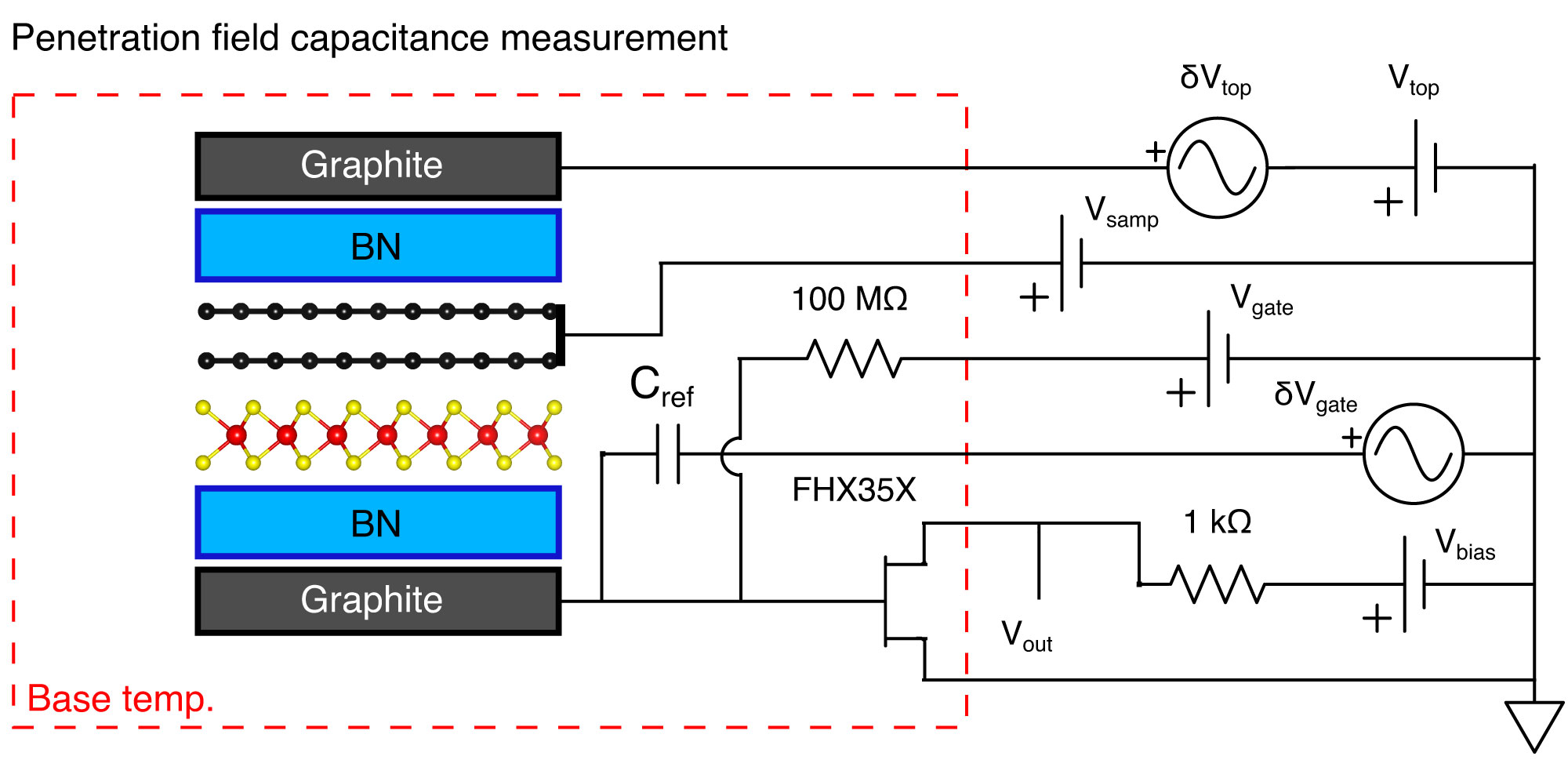}
\caption{
Electrical schematic showing the details of the penetration field capacitance measurements. The components enclosed in the red dashed box are inside the cryostat, held at base temperature. Voltages are applied to $V_{top}$ and $V_{samp}$ (at a fixed $V_{gate}$) in order to adjust charge density $n=c_Tv_T+c_Bv_B$ and displacement field $D=(c_T v_T-c_B v_B)/(2\epsilon_0)$.
}
\label{figS1b}
\end{figure*}

\begin{figure*}[t!]
\includegraphics[width=\columnwidth]{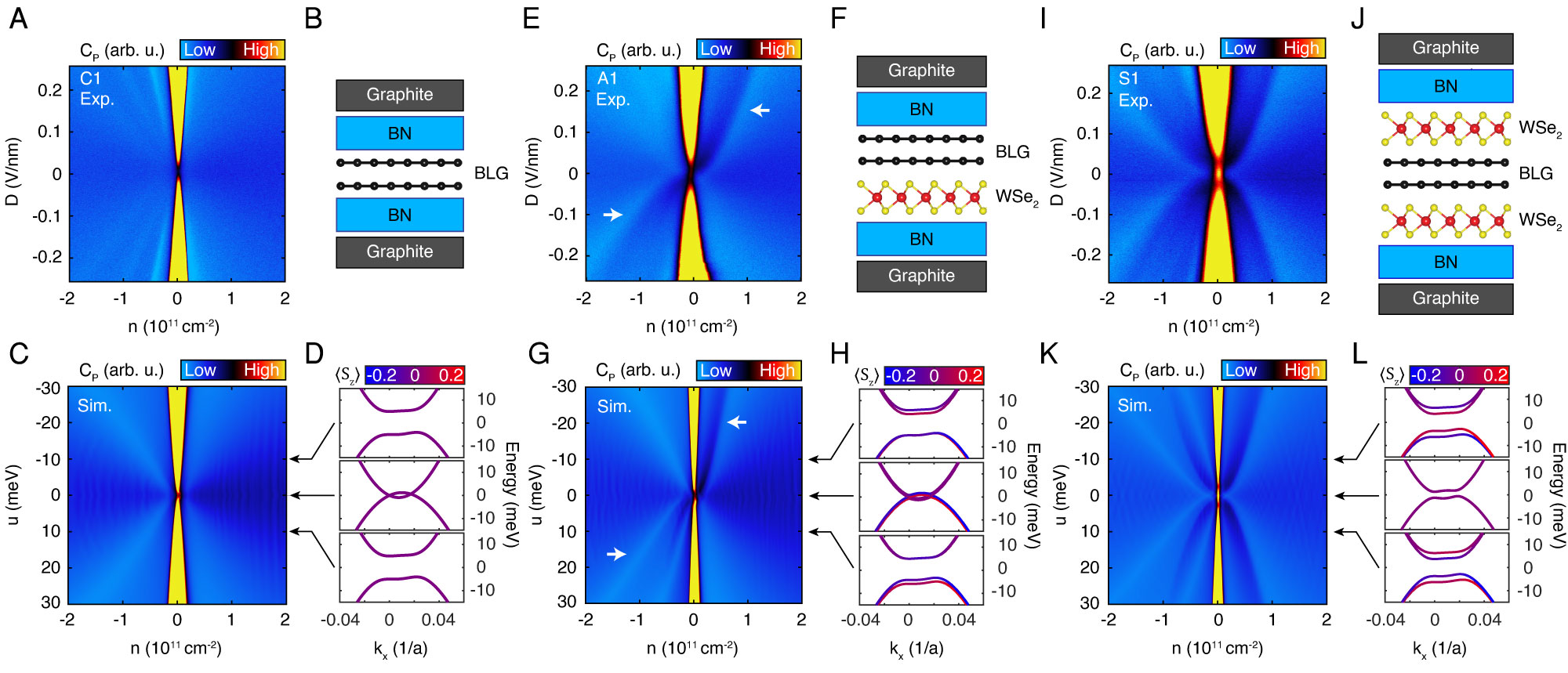}
\caption{
(A) Penetration field capacitance, $C_P$, as a function of charge density $n$ and displacement field $D$ measured at B=0 and T$\approx$50 mK in the control device C1.
(B) Schematic of the device C1, a BLG flake encapsulated with hBN.
(C) Simulated $C_P$ as a function of interlayer bias, $u$, and charge density, $n$, from a low energy continuum model for the control Device C1.
(D) Low energy bands near the K point of the Brillouin zone with $k_y=0$. Line color represents the expectation value of the out-of-plane projection of the electron spin, $\langle S_z\rangle$. Panels correspond to $u=10$ meV (top), $u=0$ meV (middle), and $u=-10$ meV (bottom).
(E) $C_P$ for device A1. Arrows indicate weak features in $C_P$.
(F) Schematic of the device A1, in which the BLG is asymmetrically encapsulated between WSe$_2$ and hBN crystals.
(G) Simulated $C_P$ from for the asymmetric geometry with $\lambda_\text{I}=1.7$~meV Ising SOC on the bottom layer. Arrows denote band-edge singularity-associated features arising from spin-split valence (conduction) bands for electron (hole) doping, visible in (H) the low energy band structure.
(I) $C_P$ measured for device S1. Note the incompressible phase centered at $D=0$, $n=0$, absent in either control or symmetric devices.
(J) Schematic of device S1, in which the BLG is symmetrically encapsulated between two few-layer WSe$_2$ crystals.
(K) Simulated $C_P$ for the symmetric geometry, with an Ising SOC of equal magnitude ($\lambda_\text{I}=2.6$ meV) but opposite signs on opposite layers.
(L) Low energy bands in the symmetric geometry near the K point of the Brillouin zone with $k_y=0$. Line color represents the expectation value of the out-of-plane projection of the electron spin, $\langle S_z\rangle$. Panels correspond to $u=10$ meV (top), $u=0$ meV (middle), and $u=-10$ meV (bottom).
}
\label{figs1d}
\end{figure*}

Small changes in device capacitance are measured using a low temperature capacitance bridge\cite{ashoori_single-electron_1992} which effectively disconnects the device capacitance from the large capacitance of the cryostat cabling, see Fig. \ref{figS1b}. $C_P$ is a measure of the capacitance between the top and bottom gates and its magnitude is high when the bilayer graphene is incompressible (gapped) and low when it is compressible (conducting). $C_P$ is measured by applying a fixed AC excitation (17-33 kHz) to the top gate ($\delta V_{top}$) and the phase and amplitude of a second AC excitation with the same frequency is adjusted and applied to a standard reference capacitor ($C_{\text{ref}}$) on the low temperature amplifier in order to balance the capacitance bridge. A commercial high electron mobility transistor (FHX35X) transforms the small sample impedance to a 1 k$\Omega$ output impedance roughly translating to a (power) gain of $\sim$1000. $V_{top}$ and $V_{samp}$ (at a fixed $V_{gate}$) are swept in order to adjust charge density $n=c_Tv_T+c_Bv_B$ and displacement field $D=(c_T v_T-c_B v_B)/(2\epsilon_0)$.

All measurements were performed within the electronic band-gap of WSe$_2$. Charge accumulation in the WSe$_2$ layers is evident in capacitance measurements at high densities and manifests as apparent negative signals in $C_P$ as charge carriers are transferred from bilayer graphene to the opposite facet of the WSe$_2$ substrates.

\clearpage

\subsection{Comparison of experimental data and numerical simulations at $B=0$ for varying $\lambda_\text{R}$}

In order to determine what type of spin-orbit symmetry breaking terms are present in the symmetric device, we analyze the band structure and simulated capacitance $C_P$ for a one sided device in further detail. As discussed in the main text, the band structure and in turn the system for the symmetric device are almost completely insensitive to the value of the Rashba SOC under the assumption of equal and opposite Rashba coupling in the top and bottom layer. To circumvent this peculiar
property we focus on the one-sided device, which does not exhibit similar insensitivity to Rashba SOC due to its asymmetric construction.

\begin{figure*}[ht]
\includegraphics[width=\linewidth]{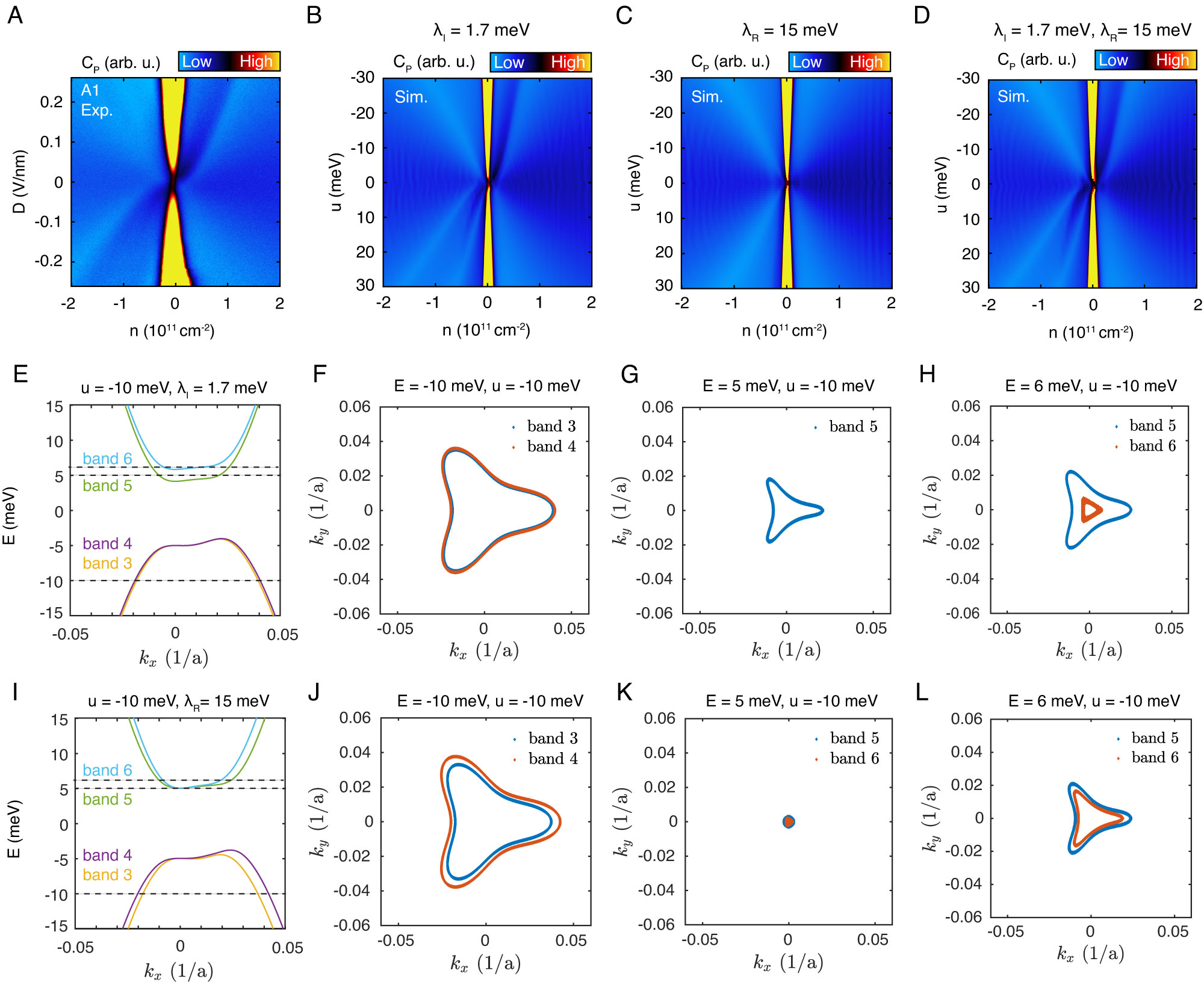}
\caption{
(A) $C_P$ measurement of device A1 as a function of $n$ and $D$.
(B) Simulated $C_P$ from a low energy continuum model with a one-sided Ising SOC of $\lambda_\text{I}=1.7$ meV.
(C) Simulated $C_P$ from a low energy continuum model with a one-sided Rashba SOC of $\lambda_\text{R}=15$ meV.
(D) Simulated $C_P$ from a low energy continuum model with a one-sided Ising and Rashba SOC of $\lambda_\text{I}=1.7$ meV, $\lambda_\text{R}=15$ meV, respectively.
(E) Low energy bands near the K point of the Brillouin zone with $k_y=0$, $u=-10$ meV, and $\lambda_\text{I}=1.7$ meV. A clear band splitting is observed in the conduction band associated with the addition of an Ising SOC.
(F) Fermi contour at $E=-10$ meV and $u=10$ meV.
(G) Fermi contour at $E=5$ meV and $u=10$ meV.
(H) Fermi contour at $E=10$ meV and $u=10$ meV.
(I) Low energy bands near the K point of the Brillouin zone with $k_y=0$, $u=-10$ meV, and $\lambda_\text{R}=15$ meV.
(J) Fermi contour at $E=-10$ meV and $u=-10$ meV.
(K) Fermi contour at $E=5$ meV and $u=-10$ meV.
(L) Fermi contour at $E=10$ meV and $u=-10$ meV.
}
\label{figS2_3}
\end{figure*}

The measured capacitance for the one sided device is shown in Fig. \ref{figS2_3}A. In addition to the symmetric gapped regions due to the applied interlayer potential (as also seen in the control device, C1), we observe two clearly defined asymmetric features (indicated with white arrows on the Fig. \ref{figS2_3}) present in the $u>0, n<0$ and $u<0, n>0$ regions. Each feature consists of two line-like ``dips" in capacitance. By definition, minima in $C_P$ correspond to maxima of density of compressibility. A maximum of compressibility in turn suggests an extremum of the band structure (van Hove singularities). With this understanding in mind we consider three separate device simulations
\begin{enumerate}
\item a pure Ising SOC system: a non-zero Ising coupling, zero Rashba coupling (Fig. \ref{figS2_3}B),
\item a pure Rashba SOC system: a zero Ising coupling, non-zero Rashba coupling (Fig. \ref{figS2_3}C),
\item a mixed system: a non-zero Ising coupling, non-zero Rashba coupling (Fig. \ref{figS2_3}D).
\end{enumerate}
Corresponding band structures for the pure Ising (Fig. \ref{figS2_3}E-H) and pure Rashba (Fig. \ref{figS2_3}I-L) devices are shown in the the panels below the capacitance simulations. We clearly see that neither Rashba nor Ising term lead to formation of a local minimum of the bandstructure. The Ising SOC together with interlayer potential causes an energy splitting of either conduction or valence bands (Fig. \ref{figS2_3}E), whilst the Rashba SOC causes primarily splittings in momenta (Fig. \ref{figS2_3}I). Of these two effects only the splitting in energy will lead to two van Hove singularities as seen in the experimental map Fig. \ref{figS2_3}A. This is further exemplified by the energy contours plots of the two systems (Fig. \ref{figS2_3}F-H and Fig. \ref{figS2_3}J-L), which show a clear difference between bandstructures for the pure Ising and pure Rashba devices. We note however that a mixed system (both a non-zero Ising and Rashba term) cannot be excluded based on this analysis as the capacitance of the mixed system possesses all qualitative features of a pure Ising device as shown in Fig. \ref{figS2_3}D.

\clearpage

\subsection{$\nu=\pm3$ phase transitions in the control device C1 and comparison with A1}

\begin{figure*}[ht]
\includegraphics[width=4.5in]{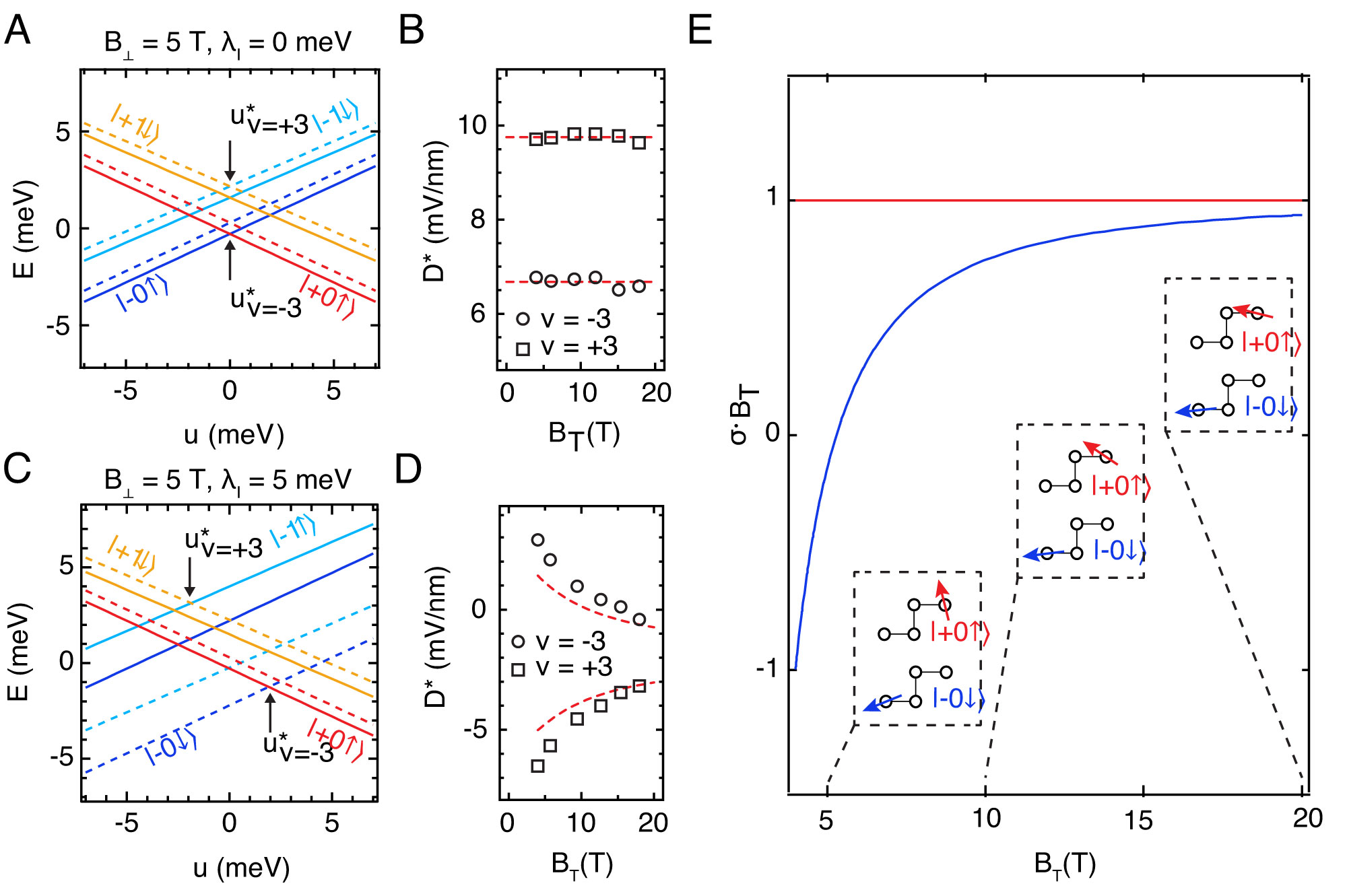}
\caption{(A) Energy level diagram of the zero-energy LL in the absence of SOC.
The $\nu=\pm3$ transitions occurring between ground states with identical spin polarization. Note that offsets from $u^*=0$ are possible due to differing on-site energies within the BLG unit cell, which can arise from coupling to the hBN substrate, but that these offsets do not influence the spin degree of freedom.
(B) Measured $D^*_{\nu=\pm3}$ as a function of $B_T$ for fixed $B_\perp=4$T in control device C1. No Zeeman dependence is observed, consistent with expectations from a SOC-free model.
(C) Energy level diagram of the zero-energy LL with a layer-selective Ising SOC of $\lambda_\text{I}=5$meV, with sign chosen so that the effect of the SOC opposes the external field (reproduced from Fig. 2F of main text). Note that the $\nu=\pm3$ transitions now occur between ground states with opposite spin polarization.
(D) Measured $D^*_{\nu=\pm3}$ as a function of $B_T$ for fixed $B_\perp=4$T in device A1, reproduced from the main text. The red dashed line is a two parameter fit with $\lambda_\text{I}=1.7$~meV and $\epsilon_{\text{BLG}}=2.8$, with the latter needed for the conversion between experimentally measured $D$ and theoretically calculated $u$.
(E) Schematic of the effect of $B_T$ in an asymmetric device. While the LL in the unaffected layer always aligns its spin polarization with the external magnetic field, the spin polarization of LLs in the SOC-proximitized result from a competition between SOC-induced Zeeman field (out of plane) and the changing direction of the physical Zeeman field. The affected spin cants only slightly for $E_Z\ll\lambda_\text{I}$, but eventually the Zeeman energy overwhelms the SOC and the two spins align as $E_Z/\lambda_\text{I}\rightarrow \infty$.}
\label{figS_LLs}
\end{figure*}

In the control device, C1, we do not expect to observe a Zeeman dependence of $D^*_{\nu=\pm3}$ as the transitions occur between states with the same spin orientation. Fig. \ref{figS_LLs}A shows the Landau level spectrum at $B=5$ T for bilayer graphene without SOC. The level transitions at $u^*_{\nu=-3}$ are between $|-0\uparrow\rangle$ and $|+0\uparrow\rangle$. The applied magnetic field simply moves these two states down in energy, shifting the energy at which the transition occurs but keeping it pinned to $u=0$ meV. This can be readily seen for the control device, C1, in the extracted $D^*$ from measurements of $C_P$, see Fig. \ref{figS_LLs}B. $D^*$ is constant across the measurable field range. Note that offsets from $u^*=0$ are possible due to differing on-site energies within the BLG unit cell, which can arise from coupling to the hBN substrate, but that these offsets do not influence the spin degree of freedom. This is in contrast with the asymmetric device with a layer specific SOC. Fig. \ref{figS_LLs}C shows the calculated Landau level spectrum for bilayer graphene now with an Ising SOC of $\lambda_\text{I}=5$ meV. The level transitions for both $\nu=-3$ and $\nu=+3$ have been shifted away from $u=0$ meV as a result of the rearrangement of the states on the bottom layer. Not only has the transition shifted from zero but the transitions now occur between states with opposite spin orientation. Application of a magnetic field acts to shift the levels in opposite directions, thereby changing the $u^*$ at which the transition occurs. This is again readily observed in the data for an asymmetric device, A1, shown in Fig. \ref{figS_LLs}D. $D ^*_{\nu=\pm3}$ moves to lower $D$ larger magnetic fields. An illustration of the canting of the spin in the spin orbit coupled layer in an asymmetric device is shown in Fig. \ref{figS_LLs}E. As the in-plane field is increased the spin in the proximitized layer cants toward the total magnetic field vector.

\clearpage

\subsection{High field $C_P$ measurements, fractional quantum hall and Chern insulator states}

\begin{figure*}[ht]
\includegraphics[width=\linewidth]{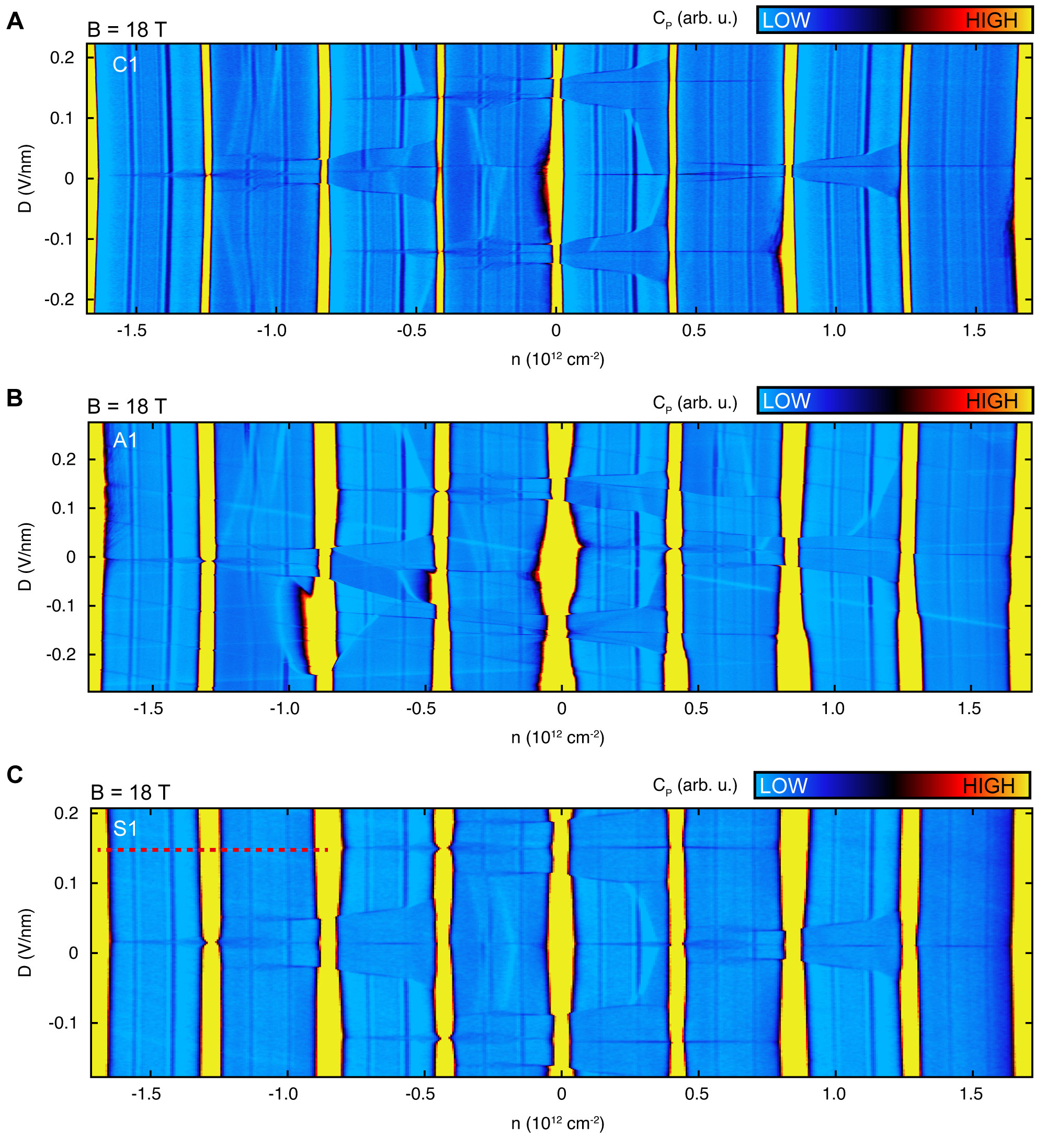}
\caption{
(A) $C_P$ for device C1 as a function of $D$ and $n$ at $B=18$ T.
(B) $C_P$ for device A1 as a function of $D$ and $n$ at $B=18$ T.
(C) $C_P$ for device S1 as a function of $D$ and $n$ at $B=18$ T.
}
\label{figS5}
\end{figure*}

\begin{figure*}[ht]
\includegraphics[width=\linewidth]{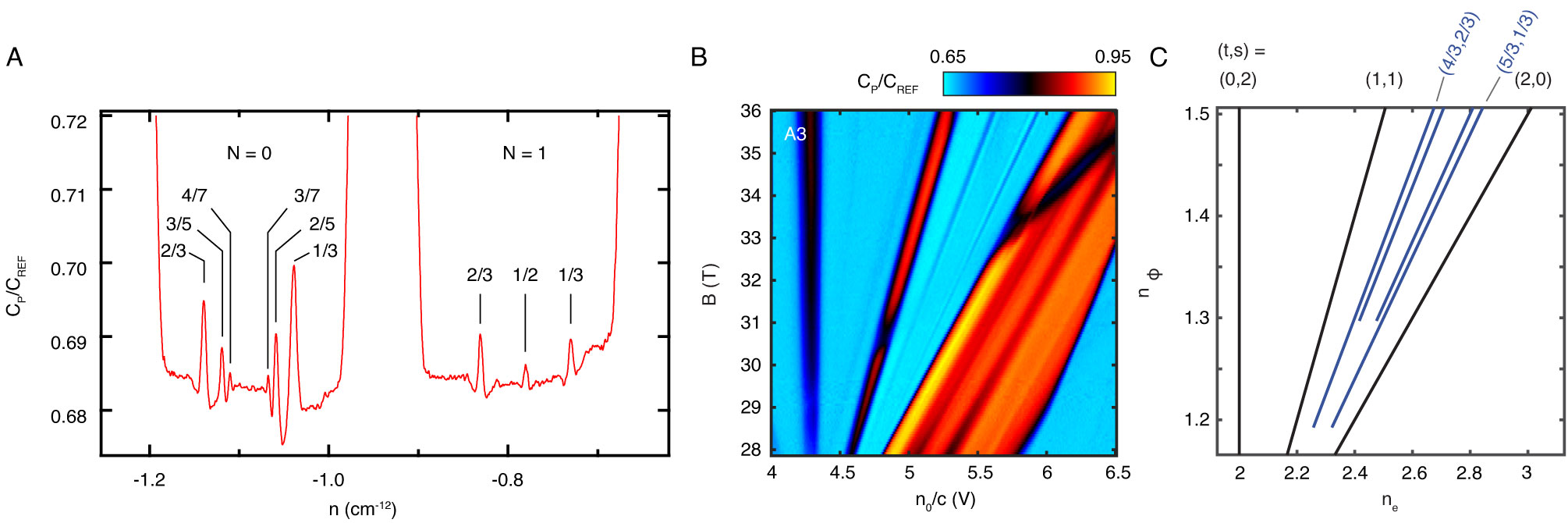}
\caption{
(A) Fractional quantum hall (FQH) states observed at 18 T in device S1. The linecut is taken at $D=1.5$ V/nm in Fig. \ref{figS5}C (red dashed line) corresponding to a range of $-4<\nu<-2$. In the N=0 orbital FQH states up to sevenths are clearly observed. In the N=1 orbital an incompressible state is observed at half filling.
(B) Fractional Chern insulator (FCI) states in asymmetric device, A3, at high magnetic fields with the BLG and hBN perfectly aligned. $C_P$ as a function of nominal electron density $n_0/c$, where $c$ is the geometric capacitance, and applied perpendicular magnetic field ($B$) at a fixed perpendicular electric field ($\frac{p_0}{c}$ = $\frac{2\epsilon_0}{c}D$ = -6 V).
(C) Schematic of the observed insulating states in units normalized to the moir\'e unit cell area ($A_{\textrm{moir\'e}}$), the number of flux quanta per moir\'e unit cell $n_{\Phi} = B A_{\textrm{moir\'e}} / \Phi_0$ and number of electrons per unit cell $n_e = n/A_{\textrm{moir\'e}}$ where $\Phi_0 = h/e$ is a flux quantum and $n$ is the electron density. The insulating states are characterized by their inverse slope and intercept in these units, $t$ and $s$, respectively. We observe a topological Chern band with $\delta t = C=1$ and $\delta s = 1$ which originates at $n_{\Phi} = 1$ between insulating states (t,s) = (1,1) and (2,0) (black lines). We observe fractional Chern insulating states at 1/3, 2/5, 3/5, 2/3rds filling of the band with quantum numbers $t,s$ = $(4/3, 2/3), (7/5, 3/5), (8/5, 2/5), (5/3, 1/3)$, respectively.
}
\label{figS4}
\end{figure*}

At higher magnetic fields we observe fractional quantum hall and Chern insulator states which are a testament to the quality of the heterostructures even with the incorporation of WSe$_2$. Fig. \ref{figS5} shows $C_P$ measurements of the zero-energy Landau level for devices C1(Fig. \ref{figS5}A), A1(Fig. \ref{figS5}B), and S1(Fig. \ref{figS5}C) taken at 18 T. In the control device, incompressible states are observed at integer and fractional fillings consistent with our previous findings\cite{zibrov_tunable_2017}. Remarkably, the the same is true for A1 and S1 where the same filling sequences are observed. The red dashed line in Fig. \ref{figS5}C shows the location of the high resolution $C_P$ linecut shown in Fig. \ref{figS4}A where fractional states are clearly observed. In device A3, fabricated with commercially obtained WSe$_2$, we observe fractional Chern insulator states at even higher fields\cite{spanton_observation_2018}. Fig. \ref{figS4}B shows $C_P$ as a function of nominal charge density $n_0/c$, where $c$ is the geometric capacitance, and $B_\perp$. Fig. \ref{figS4}C shows a schematic of the insulating states observed in Fig. \ref{figS4}B. Fractional Chern insulating states at 1/3, 2/5, 3/5, 2/3rds filling are observed within the Chern band defined between (t,s) = (1,1) and (2,0) (black lines).

\clearpage

\subsection{Asymmetries in $\nu\neq\pm3$ LL crossings}
\begin{figure*}[ht]
\includegraphics[width=\linewidth]{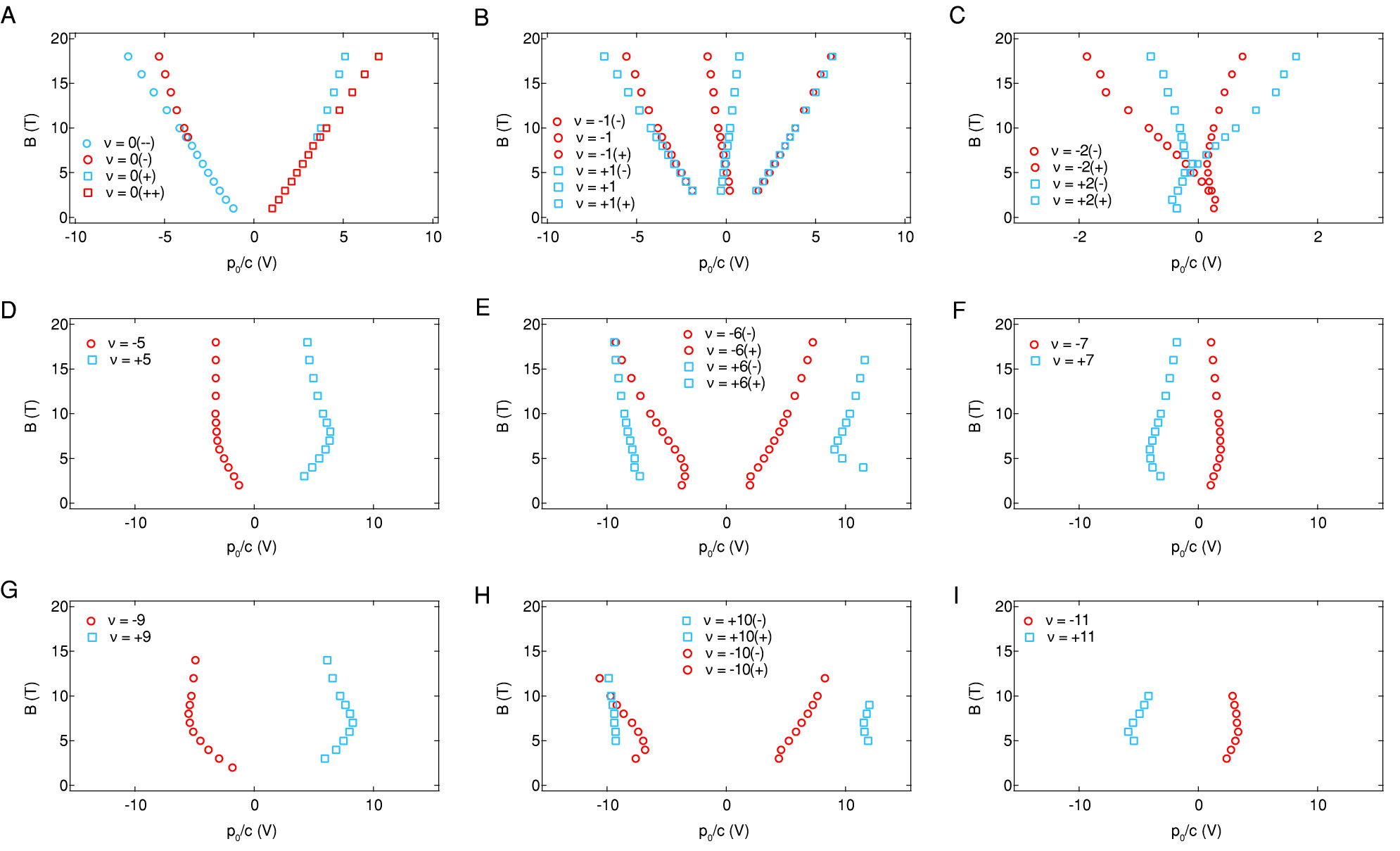}
\caption{
Level transitions in device A1 for $\nu=0$(A), $\nu=\pm1$(B), $\nu=\pm2$(C), $\nu=\pm5$(D), $\nu=\pm6$(E), $\nu=\pm7$(F), $\nu=\pm9$(G), $\nu=\pm10$(H), $\nu=\pm11$(I).
}
\label{figS7}
\end{figure*}

In addition to the crossing observed between Landau level coincidences for $\nu=\pm3$ for device A1, we also observe similar crossings at the same critical magnetic field, defined by the strength of $\lambda_\text{I}$, for $\nu=\pm1$ (Fig. \ref{figS7}B) and $\nu=\pm2$ (Fig. \ref{figS7}C). While the dependence of these crossings on magnetic field is similar to the $\nu=\pm3$ case, interactions become important for these level crossings and their full evolution is outside the scope of our single particle theoretical model. The excited state Landau levels ($N=\pm1$ Fig. \ref{figS7}D-E) and ($N=\pm2$ Fig. \ref{figS7}G-I) additionally show strong asymmetries away from zero displacement field and nonlinear dependence on magnetic field due to the proximity induced SOC.

\clearpage
\subsection{High in-plane magnetic field response and data from device A2/S2}

\begin{figure*}[ht]
\includegraphics[width=4.75 in]{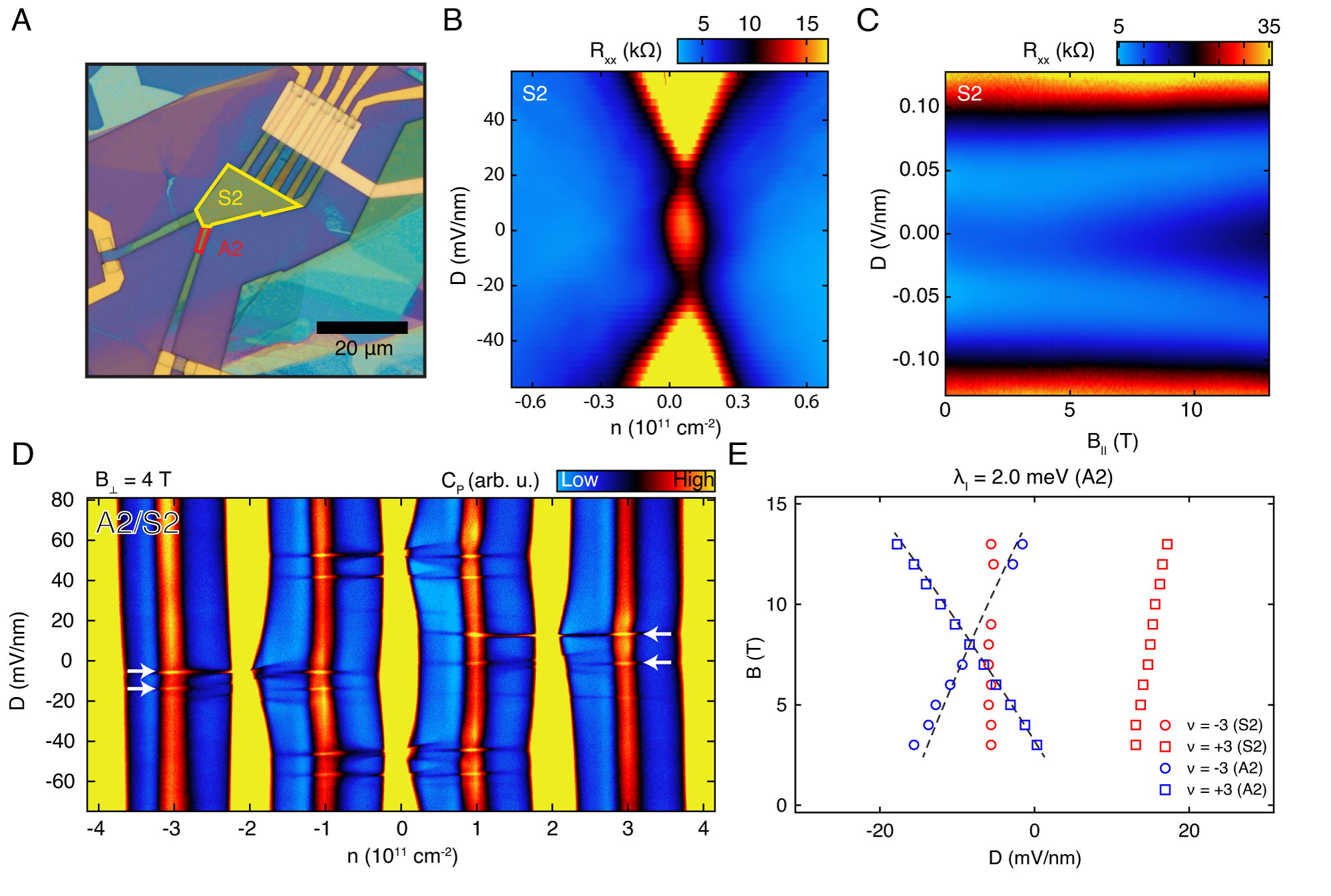}
\caption{
(A) Optical image of device A2/S2.
(B) $R_{xx}$ as a function of $D$ and $n$ at $B=0$ T for device S2. The IP is evident at charge neutrality and zero displacement field.
(C) $R_{xx}$ as a function of $B_\parallel$ and $D$ for device S2.
(D) $C_P$ as a function of $n$ and $D$ at $B=4$ T for device A2/S2. Two sets of $\nu=\pm3$ transitions are evident indicated by the white arrows.
(E) $\nu=\pm3$ transitions for device A2/S2. The crossing between $\nu=-3$ and $\nu=+3$ coming from the one-sided portion of the device (A2) is consistent with the crossing found in the asymmetric device A1. No crossing is evident in the symmetric portion which is consistent with transitions in S1 (not shown).
}
\label{figS_A2S2}
\end{figure*}

Device A2/S2 (Fig. \ref{fig3}A and Fig. \ref{figS_A2S2}) was fabricated with the intention of producing a symmetric device. However, capacitance measurements and post-facto optical microscopy show a misalignment in one region (indicated in Fig. \ref{figS_A2S2}A, resulting in a small area of the device in which only one facet of the BLG is in contact with WSe$_2$. Capacitance measurements, which are sensitive to areal averages of density of states, indeed detect twice the normal number of phase transitions (Fig. \ref{figS_A2S2}D), which show features characteristic, respectively, of both symmetric and asymmetric devices S1 and A1. In particular, one set of phase transitions shows the characteristic crossing of the $D^*_{\nu\pm3}$ transitions in finite field, resulting in a measured $\lambda_\text{I}=2.0$ meV (Fig. \ref{figS_A2S2}E). Transport measurements are performed on the side of the device that is completely encapsulated which shows clearly the inverted phase at zero field (Fig. \ref{figS_A2S2}B) and the high in-plane field response (Fig. \ref{figS_A2S2}C) that was determined in device S1 with magnetocapacitance measurements.

\newpage

\subsection{Simulations of $C_P$ at $B_\perp=0$}

In this section we describe briefly the procedure used to simulate the bulk measurements of the capacitance $C_p$.  As discussed in the main text, capacitance $C_P$ is a probe of density of states. This is expressed through the relation,
\be
C_p(\mu, u) = \frac{c}{c+\nu(\mu,u)}\,,
\ee
where we introduced a notation $\nu(\mu,u)$ for the density of states of the bilayer system at a chemical potential $\mu$ with an interlayer potential $u$ applied between the two layers. Here $c$ is a sample dependent geometric capacitance as described in the main text.

To compute the density of states $\nu(\mu,u)$, for small values of chemical and interlayer potential, we use an effective continuum model valid near the Dirac points (the $K_{\pm}$ valleys), which is expressed in the $(A1, B1, A2, B2)$ basis as \cite{hunt_direct_2017, jung_accurate_2014, mccann_electronic_2013}
\begin{eqnarray}\label{eq.HBLG0}
&&\mathcal{H}_0 = \left( \begin{array}{cccc}
\frac{u}{2} & v_0 \pi ^{\dagger} & -v_4 \pi ^{\dagger} & -v_3 \pi \\
v_0 \pi & \frac{u}{2} + \Delta ' & \gamma _1 & -v_4 \pi ^{\dagger} \\
-v_4 \pi & \gamma _1 & -\frac{u}{2} + \Delta ' & v_0 \pi ^{\dagger} \\
-v_3\pi^\dagger & -v_4 \pi & v_0 \pi & -\frac{u}{2}
\end{array} \right), \\
&&\pi = \hbar (\xi k_x + i k_y), \pi ^{\dagger} = \hbar (\xi k_x - i k_y),\quad v_{i} = \frac{\sqrt{3}a}{2\hbar} \gamma _{i}. \nonumber
\end{eqnarray}
Here, $a = 2.46\,\text{\AA}$ is the monolayer graphene lattice constant, the sign factor $\xi = \pm 1$ serves as the valley index corresponding to the valley wave vectors $\vec{K}_{\pm} = (\pm \frac{4\pi}{3a},0)$. The wavevector $\vec{k} = (k_x,k_y)$ is measured relative to $\vec{K}_{\pm}$. The hopping parameters are denoted by: $\gamma _0 = 2.61$\,eV for the interlayer nearest neighbor hopping, $\gamma _1 = 0.361$\,eV for the interlayer coupling between orbitals on the dimer sites B1 and A2, $\gamma_3 = 0.283$\,eV for the trigonal warping, and $\gamma _4 = 0.138$\,eV for the interlayer coupling between dimer and non-dimer orbitals A1 and A2 or B1 and B2. The parameter $\Delta ' = 0.015$\,eV describes the energy difference between dimer and non-dimer sites. The interlayer bias is given by $u = V_2 - V_1 = -\frac{d\epsilon_0}{\epsilon_{BLG}} D$ where $V_i$ is the potential on layer $i=1,2$ and $d$, $\epsilon_0,\epsilon_{BLG}, D$ were defined in the main text.

We include the additional symmetry breaking terms, the interfacially-induced Ising and Rashba SOC, by adding layer-specific spin-orbit Hamiltonians $\delta\mathcal{H}_i$\cite{khoo_tunable_2018}
\be
\delta \mathcal{H}_{i} = \frac{\lambda_{i}}{2}\xi \bm{s}_z + \frac{\lambda_{R,i}}{2}\left(\xi\bm{\sigma}_x \bm{s}_y-\bm{\sigma}_y \bm{s}_x\right)
\ee
to the total Hamiltonian of the system $\mathcal{H} = \mathcal{H}_{0} + \sum_{i=1,2} \delta \mathcal{H}_{i}$. Here $i=1,2$ again labels the layers, $\lambda_i$ and $\lambda_{R,i}$ are the Ising and Rashba spin-orbit couplings in the layer $i$. The Pauli matrices $\bm{\sigma}_i$ and $\bm{s}_i$ denote the sublattice and spin degrees of freedom.

The explicit calculation of the density of states is done by discretizing the momentum $\vec{k}$ on a lattice, diagonalizing the total Hamiltonian $\mathcal{H}$ and then evaluating a sum
\be
\nu(\mu) = \frac{1}{4 \pi^2} \frac{S_k}{N} \sum_{\vec{k}, j} \delta(\epsilon_{\vec{k}, j}-\mu)\,,
\ee
where $N = \sum_{\vec{k}} 1$ is the total number of momentum points in the Brillouin zone with area $S_{k}$. The index $j$ runs over all energy bands, each with single-particle energy $\epsilon_{\vec{k},j}$, and hence the valley/spin degeneracy is included explicitly in this summation. To overcome numerical artifacts due to a finite discretization of the momentum $\vec{k}$, we soften the delta function in the above summation with a Lorentzian as
\be
\delta(\epsilon_{\vec{k}, j}-\mu) \approx \frac{\gamma}{\pi} \frac{1}{(\epsilon_{\vec{k}, j}-\mu)^2+\gamma^2}\,.
\ee
For all simulations plotted in the text we chose $\gamma = 0.12$ meV, which required a $k$ spacing of $10^{-3}$ 1/$a$ to ensure lack of ringing artifacts.

In order to convert from a simulated parameter space $(\mu, u)$ to the experimentally accessible one $(n(\mu,u), u)$ we evaluate the charge density through a relation\cite{hunt_direct_2017}
\be
e n(\mu,u) = c\mu + \int_{0}^{\mu} d\tilde{\mu} ~\nu(\tilde{\mu}, u)
\ee
and then plot the capacitance $C_p$ as a function of charge density $n(\mu,u)$ and interlayer potential $u$ as shown in Fig. 1 in the main text and the supplemental figures.

This analysis readily extends itself to include an in-plane magnetic field $B_\parallel$ as done in Fig. 3 in the main text. We incorporate the field dependence only via the Zeeman splitting term (without loss of generality we take $B_\parallel$ along the $x$-axis)
\be\label{eq.Binplane}
\mathcal{H}_{B_\parallel} = \mu_B B_\parallel \bm{s}_x
\ee
where $\mu_B = 5.796\times10^{-5}$ eV/T. A system with magnetic field is then characterized by a Hamiltonian $\mathcal{H} = \mathcal{H}_{0} + \sum_{i=1,2} \delta\mathcal{H}_{i} + \mathcal{H}_{B_\parallel}$ and the procedure followed to obtain the capacitance maps in in-plane magnetic field is then identical to the one without magnetic field applied.

\subsection{Landau Level Calculations}

Here we show how the zero Landau level (ZLL) spectra are calculated for one-sided devices A1-A3. The approach is detailed in Ref. \cite{khoo_tunable_2018} and in this section we briefly summarize the procedure and elaborate on additional modifications required to produce experimental fits.

In the absence of magnetic field, the Hamiltonian of a one-sided device is given by $\mathcal{H} = \mathcal{H}_{0} + \delta\mathcal{H}_{1}$. This Hamiltonian can be extended to the case of large perpendicular magnetic field by introducing spin-resolved Landau level creation and annihilation operators, as well as introducing the Zeeman splitting term in the perpendicular direction,
\be
\mathcal{H}_{B_\perp} = \mu _B B_{\perp} \bm{s}_z.
\ee
When $\gamma_3 = 0$, by careful construction of the ansatz, the system decouples into subspaces labelled by the valley index $\xi$, as well as the largest Landau level index present within each subspace $n$. The Hamiltonian of the full system can therefore be written as a direct sum of the subspace-specific Hamiltonians, $\mathcal{H} = \bigoplus_{\xi,n}\mathcal{H}_{\xi,n}$. The ZLLs are then given by the eight smallest eigenvalues of $\mathcal{H}$, which comprises the eigenvalues of $H_{\xi,0}$ (two), the two smallest eigenvalues of $H_{\xi,1}$ (four), and the smallest eigenvalue of $H_{\xi,2}$ (two).

In this work, we introduce two additional modifications to fit experimental data as follows:
\begin{enumerate}
\item
\textit{Sublattice asymmetry}. This is due to on-site potential differences between the four different sublattice sites of BLG, which we model by adding to $\mathcal{H}$ the layer specific Hamiltonians
\be
\mathcal{H}^{AB}_{i} = \frac{\Delta_{AB,i}}{2}\bm{\sigma}_z.
\ee
Once again $i$ labels the layers. These terms appear in the diagonal entries so that while each of the Landau levels are shifted, no mixing is introduced.

The parameters $(\Delta_{AB,1},\Delta_{AB,2})$ used for fitting are $(0.44,0)$meV for Fig. 2(G-H) in the main text, and $(57.3,-50)$meV for Fig.S8(E) in the supplementary material.

\item
\textit{In-plane magnetic field}. This effect is included by adding the Zeeman splitting term given $\mathcal{H}_{B_\parallel}$ given in Eq. \ref{eq.Binplane}. However, it introduces mixing across all the subspaces of different $n$ but with $\xi$. Its leading corrections are captured by introducing the in-plane Zeeman term only to the components spanned by the ZLL eigenstates in the $\gamma_4 =0$ limit. These are $\left\lbrace|A1s,0\rangle, |A1s,1\rangle, |A2s,0\rangle \right\rbrace$ for $\xi = +$, and $\left\lbrace|B2s,0\rangle, |B1s,0\rangle, |B2s,1\rangle \right\rbrace$ for $\xi = -$, so that the leading corrections of the in-plane $B$-field are captured by solving for the eigenvalues of the truncated system $\mathcal{H} \simeq \bigoplus _{\xi,n=0,1,2} \mathcal{H}_{\xi,n}$.
\end{enumerate}

The values of interlayer potential $u$, or equivalently the displacement field $D$, at which the two most negative (positive) ZLLs cross can then be extracted at each given value of applied magnetic field ${\bf B} = (B_\parallel, B_\perp)$ to obtain the crossing plots for $\nu = -3 (+3)$.

Finally, we note that while the $\gamma_3$ parameter introduces mixing between the ZLLs and the higher LLs, the effect is extremely weak such that the ZLL spectrum remains essentially unchanged. We verified that this is indeed the case and therefore neglect it completely in the ZLL calculations as in previous work\cite{hunt_direct_2017}.

\subsection{Edge state calculations}
The edge spectra shown in Fig.~3 were obtained by diagonalizing a tight-binding model of a zig€-zag edge strip of width $W = 1000$ for the lowest 32 states. Tight binding parameters are $t_0 = -2.6, t_1 = 0.36, t_3 = 0.28$ (in eV) taken from the notation and valyes of Ref.~\onlinecite{jung_accurate_2014}.  We implemented SOC  using the minimal hoppings required to reproduce the continuum Ising (next-nearest neighbor hopping) and Rashba (nearest neighbor hopping)  couplings. In the insets, color denotes the spin $\langle s^z \rangle$, and data is only shown for states with position localized to $\langle \hat{y} \rangle < \frac{3}{4} W$.  The SOC was chosen to be  completely layer asymmetric (preserving the 3D inversion symmetry) with $\lambda_{\textrm{I}} = 5$meV and $\lambda_{\textrm{R}} = 15$meV. The Rashba coupling does open up an edge gap, but it is undetectable on this scale because it is suppressed by powers of $g / t_1$ where $g$ are small couplings like $\lambda_{\textrm{R}}$ and $D$. We have verified that the edge gap does grow continually as  $\lambda_{R}$ is increased to physically unrealistic values.

Note that these spectra neglect the particle-hole breaking hopping $t_4 = \sim 0.14$eV. This term changes neither the connectivity of the edge states nor the magnitude of their gaps, but it does alter their dispersion. In the a tight-binding model we use here, in which hoppings are abruptly terminated at the edge, the resulting edge dispersion has a Fermi surface with compensated electron and hole pockets. It is not clear whether these would persist in more realistic models, a useful direction for future ab-initio calculations.

\end{document}